\documentclass[journal=jceaax,manuscript=article]{achemso}

\usepackage{chemformula} 
\usepackage{}
\usepackage{bm}
\usepackage[T1]{fontenc} 
\usepackage{subcaption}
\usepackage{amssymb}
\usepackage{multirow}
\usepackage{longtable}
\usepackage{rotating}
\usepackage{hhline}
\usepackage{amsmath}
\usepackage{graphicx}
\usepackage{color}
\usepackage{tikz}
\usetikzlibrary{shapes.geometric}
\SectionNumbersOn
\usepackage{lineno}

\author{Nicolas~Hayer}
\affiliation{Laboratory of Engineering Thermodynamics, RPTU Kaiserslautern, Erwin-Schrödinger-Str. 44, 67663 Kaiserslautern, Germany}

\author{Hans~Hasse}
\affiliation{Laboratory of Engineering Thermodynamics, RPTU Kaiserslautern, Erwin-Schrödinger-Str. 44, 67663 Kaiserslautern, Germany}

\author{Fabian~Jirasek\textsuperscript}
\affiliation{Laboratory of Engineering Thermodynamics, RPTU Kaiserslautern, Erwin-Schrödinger-Str. 44, 67663 Kaiserslautern, Germany}
\email{fabian.jirasek@rptu.de}

\title{Modified UNIFAC 2.0 -- A Group-Contribution Method Completed with Machine Learning}

\begin{document}
%
%
%
%
%


\begin{abstract}
Predicting thermodynamic properties of mixtures is a cornerstone of chemical engineering, yet conventional group-contribution (GC) methods like modified UNIFAC (Dortmund) remain limited by incomplete tables of pair-interaction parameters. To address this, we present modified UNIFAC 2.0, a hybrid model that integrates a matrix completion method from machine learning into the GC framework, allowing for the simultaneous training of all pair-interaction parameters, including the prediction of parameters that cannot be fitted due to missing data. Utilizing an extensive training set of more than 500,000 experimental data for activity coefficients and excess enthalpies from the Dortmund Data Bank, modified UNIFAC 2.0 achieves improved accuracy compared to the latest published version of modified UNIFAC (Dortmund) while significantly expanding the predictive scope. Its flexible design allows updates with new experimental data or customizations for specific applications. The new model can easily be implemented in established simulation software with complete parameter tables readily available.
\end{abstract}

\section{Introduction}
Understanding the thermodynamic properties of mixtures is essential for chemical engineering. Due to the impracticality of studying each relevant mixture experimentally, reliable prediction methods are crucial. Group-contribution (GC) methods offer an efficient solution by decomposing molecules into structural groups, significantly reducing the number of parameters and enabling extrapolations to unstudied components and mixtures. The most successful GC method in chemical engineering is probably UNIFAC~\cite{Fredenslund.1975}, which is available in different versions~\cite{Magnussen.1981,Wittig.2003,UNIFAC_TUC.2023,Constantinescu.2016}. UNIFAC is a model for predicting the excess Gibbs energy of mixtures and derived properties, such as activity coefficients and excess enthalpies. It has been widely adopted for describing reaction and phase equilibria in mixtures and is implemented in all relevant process simulators~\cite{DDB.2024,ChemstationsInc..2024,AspenTechnologyInc..2024}. 

However, UNIFAC has important drawbacks: Firstly, the most comprehensive versions of UNIFAC, namely, original UNIFAC~\cite{Wittig.2003} and modified UNIFAC (Dortmund)~\cite{Constantinescu.2016}, have been regularly updated, but only up to 2003~\cite{Wittig.2003} and 2016~\cite{Constantinescu.2016}, respectively. Since then, the work on UNIFAC updates has continued, but only commercially within the so-called UNIFAC-Consortium (TUC)~\cite{UNIFAC_TUC.2023}, so the latest UNIFAC versions are not publicly available. Furthermore, the applicability of all UNIFAC versions, including the commercial ones, is limited by the availability of pair-interaction parameters between structural groups. These parameters are derived from vapor-liquid equilibrium (VLE) and other thermodynamic data of mixtures, leaving substantial gaps when no suitable training data are available, severely hampering the applicability of UNIFAC.

Compared to the original UNIFAC~\cite{Wittig.2003}, in which two parameters are used to describe the interactions between a given pair of groups, modified UNIFAC~\cite{Constantinescu.2016} considers the temperature dependence of these parameters by a simple function, leading to up to six parameters that can be adjusted for a given pair of groups. This increased flexibility often improves accuracy in describing different mixtures, making modified UNIFAC (Dortmund) arguably the best GC method presently available. For simplicity, we will label the latest public version of modified UNIFAC (Dortmund), which we use as the reference here, as mod.~UNIFAC 1.0. Mod.~UNIFAC 1.0 considers 63 \textit{main groups}, subdivided into 125 \textit{subgroups}. While each subgroup $k$ has individual size parameters describing their surface area ($Q_k$) and volume ($R_k$), which are reported for all 125 defined subgroups, pair-interaction parameters are defined between main groups $m$ and $n$. In the current parameterization of mod.~UNIFAC 1.0, these interaction parameters are reported for only 39\% of all possible pairs of main groups; Fig.~S.1 in the Supporting Information illustrates this. This situation significantly hampers the applicability of mod.~UNIFAC 1.0 since a single missing group pair-interaction parameter for a given mixture prevents the use of the method.

Consequently, the pair-interaction parameters of mod.~UNIFAC 1.0, which are asymmetric ($a_{mn} \neq a_{nm}$, $b_{mn} \neq b_{nm}$, $c_{mn} \neq c_{nm}$), can be arranged in (sparsely filled) matrices, making the prediction of the missing parameters a matrix completion problem, for which matrix completion methods (MCMs) from machine learning (ML)~\cite{A.Ramlatchan.2018, Koren.2009} can be used. We have demonstrated the applicability of MCMs in thermodynamics in prior work, where we have developed MCMs to predict different thermodynamic properties of mixtures~\cite{Jirasek.2020, Jirasek.2020b, Damay.2021, Hayer.2022, Gromann.2022} and different types of pair-interaction parameters~\cite{Jirasek.2022,Jirasek.2023}. Most importantly, we have recently introduced UNIFAC 2.0~\cite{Hayer.2024}, a hybrid model that embeds an MCM into the framework of the original UNIFAC model~\cite{Wittig.2003}. Through this integration, the MCM predicts the missing pair-interaction parameters between the main groups of original UNIFAC. UNIFAC 2.0 was trained on experimental activity coefficients derived from binary VLE data and limiting activity coefficient data taken from the Dortmund Data Bank (DDB)~\cite{DDB.2023} in an end-to-end manner, avoiding the sequential and often intuitive approaches that have characterized the traditional fitting process of UNIFAC. Our recent work demonstrates that the hybrid UNIFAC 2.0, based on a learned completed pair-interaction parameter table, outperforms the original UNIFAC method in terms of scope and accuracy~\cite{Hayer.2024}.

In this work, we transfer the concept of embedding an MCM in GC methods from original UNIFAC to mod.~UNIFAC and introduce mod.~UNIFAC 2.0. Similar to UNIFAC 2.0~\cite{Hayer.2024}, mod.~UNIFAC 2.0 exhibits complete pair-interaction parameterizations and was trained end-to-end on an extensive database of more than 500,000 data points from the DDB. As the consideration of the temperature dependence of the group interactions makes mod.~UNIFAC more flexible, we have included experimental data on the excess enthalpy besides data on activity coefficients in the training process of mod.~UNIFAC 2.0. 

By retaining the mod.~UNIFAC equations, mod.~UNIFAC 2.0 maintains the high accessibility of the original model and can easily be implemented in process simulators by simply replacing the parameter sets with the ones freely provided in the Supporting Information of this work. At the same time, mod.~UNIFAC 2.0 eliminates the most significant limitation of the original model by filling all gaps in the pair-interaction parameter tables, tremendously increasing the applicability to any mixture whose components can be represented by the presently defined structural groups. The subgroup-specific size parameters $R_k$ and $Q_k$ for using mod.~UNIFAC 2.0, which are identical to those of the published mod.~UNIFAC 1.0 version, are also provided in the Supporting Information of this work.

\section{Development of Mod.~UNIFAC 2.0}
\subsection{General Framework}
\begin{figure}[H]
    \centering
    \includegraphics[width=\textwidth]{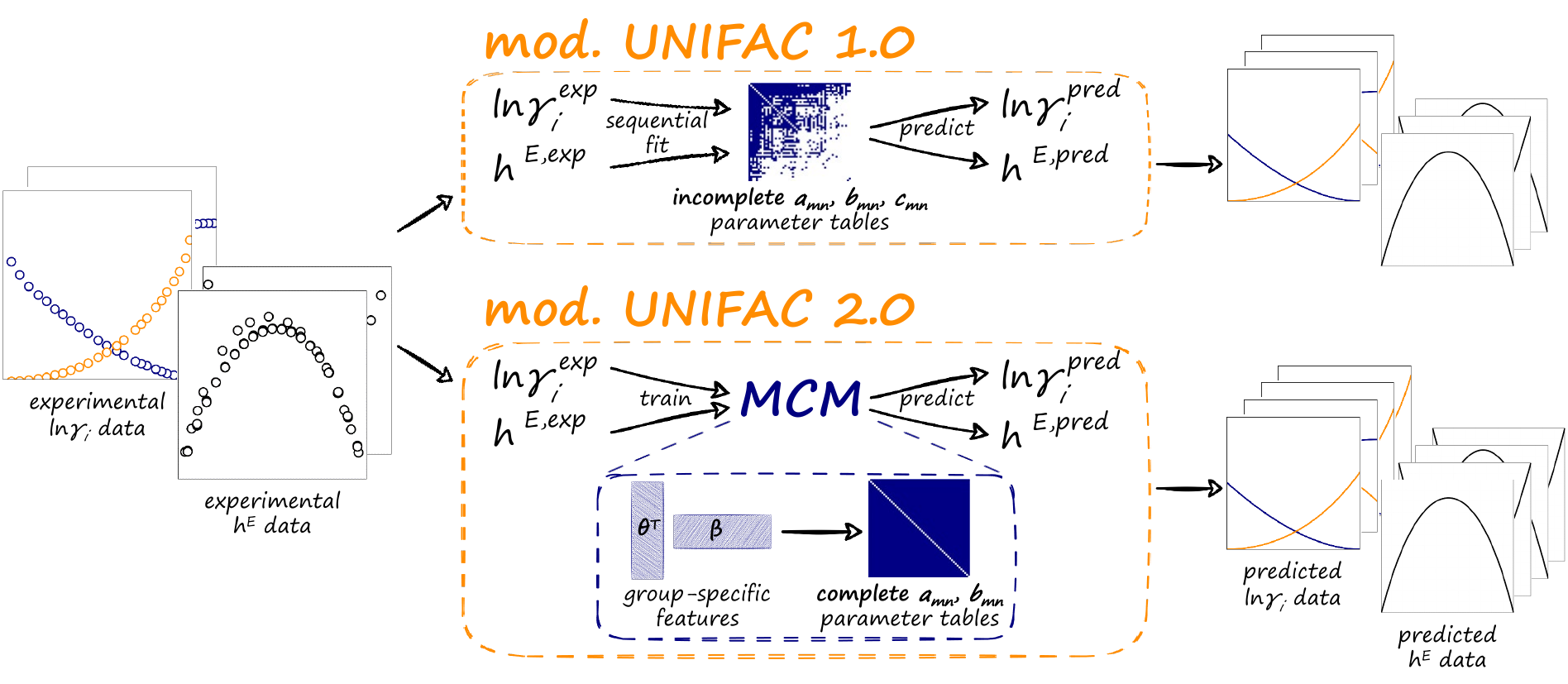}
    \caption{Comparison of mod.~UNIFAC 1.0~\cite{Constantinescu.2016} and mod.~UNIFAC 2.0 (this work). Mod.~UNIFAC 1.0 relies on sequential parameter fitting, whereas mod.~UNIFAC 2.0 integrates a matrix completion method (MCM) for predicting pair-interaction parameters into the mod.~UNIFAC framework. Mod.~UNIFAC 2.0 was trained end-to-end on experimental logarithmic activity coefficients ($\ln\gamma_i$) and excess enthalpy ($h^\text{E}$) data. After training, the completed pair-interaction parameter matrices facilitate predictions of thermodynamic properties for a vast range of binary and multi-component mixtures.}
    \label{modUNI20_fig:Scheme}
\end{figure}

Fig.~\ref{modUNI20_fig:Scheme} illustrates how mod.~UNIFAC 2.0 was developed by embedding an MCM into the mod.~UNIFAC framework. The resulting method was trained end-to-end on experimental logarithmic activity coefficients ($\ln\gamma_i$) and excess enthalpies ($h^\text{E}$) in binary mixtures. The $\ln\gamma_i$ were obtained from the limiting activity coefficient database of the DDB and derived from binary VLE data, cf.~Section "Data" for details. Mod.~UNIFAC 2.0 is compared here to mod.~UNIFAC 1.0, which uses the same structural groups and physical model equations as mod.~UNIFAC 2.0 but whose parameters were obtained by sequential parameter fitting on a data basis that includes only data taken before 2016. Additionally, mod.~UNIFAC 1.0 was trained on additional mixture properties beyond those included here~\cite{Gmehling.2012,Constantinescu.2016}.
Mod.~UNIFAC 1.0 extends the parameter $\Psi_{nm}$ of the original UNIFAC model by introducing a temperature dependence through the additional interaction parameters $b_{mn}$ and $c_{mn}$:
\begin{equation}
    \Psi_{nm}=\exp\left({-\frac{a_{nm}+b_{nm}T+c_{nm}T^2}{T}}\right)
    \label{modUNI20_eq:modUNIFAC_Psi}
\end{equation}
Setting $b_{mn}=c_{mn}=0$ results in the original UNIFAC definition of $\Psi_{nm}$~\cite{Wittig.2003}.

However, in mod.~UNIFAC 1.0, $c_{mn}$ parameters were fitted for only very few pairs of groups, and $c_{mn}=0$ is used for most group combinations. Therefore, we have decided to only use $a_{mn}$ and $b_{mn}$ in mod.~UNIFAC 2.0, which are modeled by two MCMs trained to decompose the two matrices containing the parameters $a_{mn}$ and $b_{mn}$, respectively, into the product of two respective feature matrices. Each pair-interaction parameter is thereby modeled as:
\begin{align}
    \label{modUNI20_eq:matrix_factorization_a}
    a_{mn} &= \bm{\theta}_m^{a}\cdot\bm{\beta}_n^a \\
    \label{modUNI20_eq:matrix_factorization_b}
    b_{mn} &= \bm{\theta}_m^{b}\cdot\bm{\beta}_n^b
\end{align}
Here, $\bm{\theta}_m^a$, $\bm{\theta}_m^b$, $\bm{\beta}_n^a$, and $\bm{\beta}_n^b$ are vectors of length $K$, where $K$ is called latent dimension. This hyperparameter was determined in preliminary studies and set to $K=8$. For simplicity, we collectively refer to the feature vectors as $\theta$ and $\beta$ in the following.

All parameters of mod.~UNIFAC 2.0 are learned \textit{simultaneously}, which is in sharp contrast to the sequential approach used in the original model. We have trained mod.~UNIFAC 2.0 within a Bayesian framework, treating each experimental data point ($\ln\gamma_i$, $h^\text{E}$), feature ($\theta$, $\beta$), and interaction parameter ($a_{mn}$, $b_{mn}$) as random variables drawn from probability distributions. By applying Bayes' theorem, we link these variables through three key distributions: the prior, the likelihood, and the posterior.

The prior represents initial assumptions about the features before observing data. Here, the prior for all features is a standard normal distribution, $\mathcal{N}(0, 1)$, which is uninformative and introduces no bias toward specific feature values, except for discouraging very large values, thereby serving as a kind of regularization. This choice provides a simple and effective starting point for learning features from the empirical data.

The likelihood defines the probability of observing the data ($\ln\gamma_i^{\text{exp}}$ and $h^\text{E,exp}$) given the features. It is modeled using a Cauchy distribution centered around the predicted values $\ln\gamma_i^{\text{pred}}$ and $h^\text{E,pred}$, respectively:\begin{align}
p(\ln\gamma_i^{\text{exp}}|\theta,\beta) &= \text{Cauchy}(\ln\gamma_i^{\text{pred}}, \lambda) \\
p(h^\text{E,exp}|\theta,\beta) &= \text{Cauchy}(h^\text{E,pred}, \lambda)
\end{align}
where $\lambda$ is the scale parameter of the Cauchy distribution, which was set to $\lambda=0.4$ as in our recent work~\cite{Hayer.2024}. The Cauchy distribution's heavy tails make the model robust to outliers or flawed training data, mitigating the influence of experimental noise during training. Predicted values for $\ln\gamma_i^{\text{pred}}$ and $h^\text{E,pred}$ are obtained using the standard mod.~UNIFAC equations, which are fully described in Refs.~\cite{Weidlich.1987,Gmehling.1993}:
\begin{align}
\label{modUNI20_eq:lngamma}
\ln\gamma_i^{\text{pred}} &= \text{mod.~UNIFAC}(a_{mn}, b_{mn}, R_k, Q_k, \bm{x}, T) \\
\label{modUNI20_eq:hE}
h^\text{E,pred} &= -RT^2 \sum_{i=1}^N x_i \left( \frac{\partial \ln\gamma_i^{\text{pred}}}{\partial T} \right)_{p, \bm{x}}
\end{align}
where $\bm{x}$ is the composition vector (for binary mixtures, this reduces to $x_1$), $T$ is the temperature, and $a_{mn}$ and $b_{mn}$ are the predicted pair-interaction parameters of mod. UNIFAC 2.0 calculated from the learned features according to Eqs.~(\ref{modUNI20_eq:matrix_factorization_a}) and (\ref{modUNI20_eq:matrix_factorization_b}).

The goal of Bayesian inference is to find the posterior, which combines the prior and the likelihood, i.e., it encapsulates updated beliefs about the features after considering both prior information and empirical data. Using Pyro, a probabilistic programming language written in Python and supported by PyTorch~\cite{Bingham.2018}, we have approximated the posterior using stochastic variational inference (VI) under the mean-field assumption~\cite{Blei.2017}, where all features are considered independent, and a normal variational distribution approximates each. During this step, the evidence lower bound (ELBO) was maximized using the Adam optimizer~\cite{Kingma.22.12.2014} with a learning rate of 0.15, ensuring efficient and scalable learning over the large experimental data set.

The result of training mod.~UNIFAC 2.0 is a learned probability density for each feature, from which we used the means to calculate the final pair-interaction parameters (cf.~Eqs.~(\ref{modUNI20_eq:matrix_factorization_a}) and (\ref{modUNI20_eq:matrix_factorization_b})), which are subsequently plugged into the mod.~UNIFAC equations~\cite{Weidlich.1987,Gmehling.1993} to give predictions for unstudied activity coefficients.

We have made the complete final set of pair-interaction parameters -- derived from training mod.~UNIFAC 2.0 on the entire database (see Section "Data") -- freely available in the Supporting Information as .csv files. Additionally, we provide the subgroup-specific size parameters $R_k$ and $Q_k$, which are identical to the published mod.~UNIFAC 1.0 version~\cite{Constantinescu.2016}.

\subsection{Data}
Experimental data for activity coefficients $\gamma_{i}$ and excess enthalpies $h^\text{E}$ in binary mixtures were used for training mod.~UNIFAC 2.0. All data were taken from the most extensive database for thermodynamic properties, the DDB~\cite{DDB.2024}. During preprocessing, data points that were considered to be of low quality by the DDB were excluded. We also restricted our selection to binary mixtures whose components could be decomposed into the mod.~UNIFAC subgroups. Additionally, the VLE data were limited to pressures up to 10 bar.

After preprocessing, the $h^\text{E}$ data set comprises 259,707 data points for 8,735 binary mixtures. The data set for $\gamma_{i}$ consists of 243,257 data points for 21,452 binary mixtures, which was obtained by combining 68,642 data points for limiting activity coefficients and 174,615 data points calculated from VLE data using the extended Raoult's law assuming an ideal gas phase\footnote[1]{We acknowledge that at 10 bar, deviations from this assumption have to be expected. The 10 bar limit was chosen as a compromise between limiting these deviations and losing interesting systems from the database. We have refrained from including fugacity coefficients to correct for the non-ideality of the gas phase for computational reasons.} and neglecting the pressure dependence of the chemical potential in the liquid phase: 
\begin{equation}
    \gamma_{i}(T,\bm{x}) = \frac{p \cdot y_{i}}{p_i^{\text{s}}(T)\cdot x_{i}}
    \label{modUNI20_eq:Raoult}
\end{equation} 
Here, $p$ corresponds to the total pressure and $p_i^{\text{s}}$ to the vapor pressure of the pure component $i$, while $x_{i}$ and $y_{i}$ are the mole fractions of component $i$ in the liquid and vapor phases, respectively.

\section{Results and Discussion}
\subsection{Overall Performance of Mod.~UNIFAC 2.0}
\begin{figure}[H]
    \centering
    \includegraphics[width=0.8\textwidth]{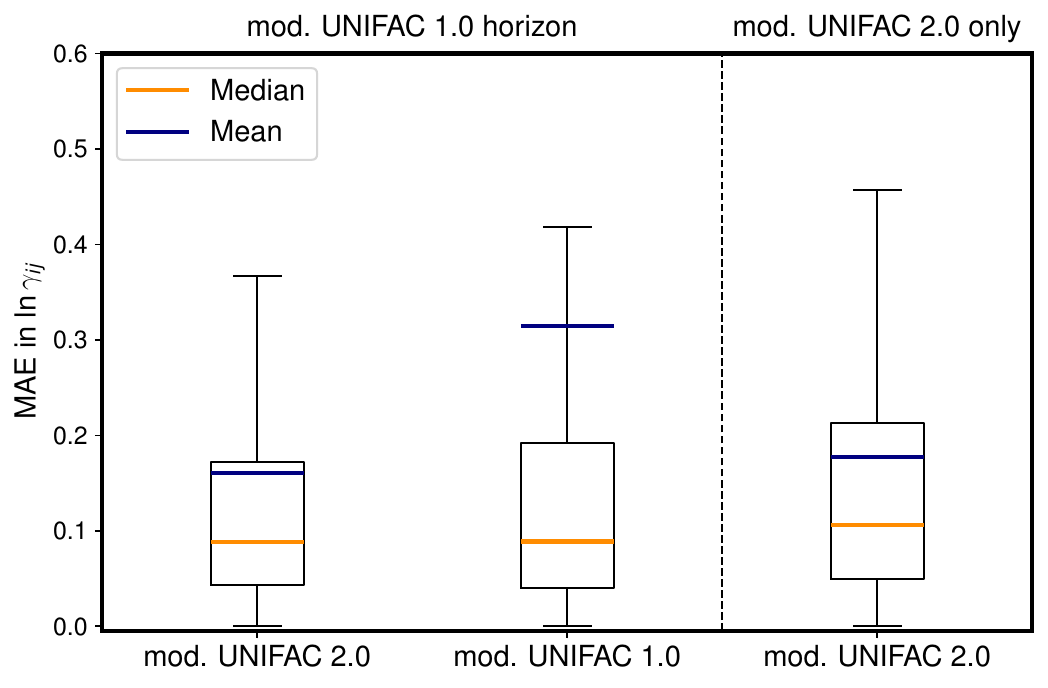}
    \caption{Mean absolute error (MAE) of the predicted $\ln\gamma_i$ with mod.~UNIFAC 2.0 and comparison to mod.~UNIFAC 1.0 for those mixtures that can also be predicted by the latter model ("mod.~UNIFAC 1.0 horizon"). The "mod.~UNIFAC 1.0 horizon" comprises 221,639 data points for 16,932 binary mixtures, while an additional 21,618 experimental data points for 4,520 binary mixtures could only be predicted with mod.~UNIFAC 2.0 ("mod.~UNIFAC 2.0 only"). The boxes represent the interquartile ranges (IQR), and the whiskers extend to the last data points within 1.5 times the IQR from the box edges.}
    \label{modUNI20_fig:Performance_lngamma_bestModel}
\end{figure}

\begin{figure}[H]
    \centering
    \includegraphics[width=0.8\textwidth]{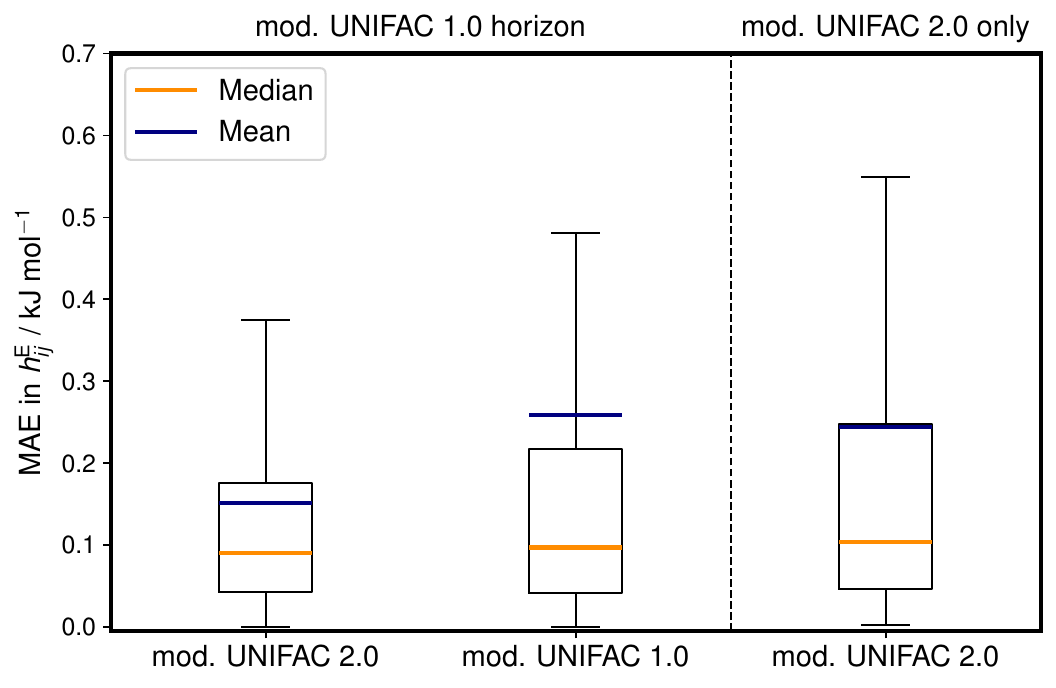}
    \caption{Mean absolute error (MAE) of the predicted $h^\text{E}$ with mod.~UNIFAC 2.0 and comparison to mod.~UNIFAC 1.0 for those mixtures that can also be predicted by the latter model ("mod.~UNIFAC 1.0 horizon"). The "mod.~UNIFAC 1.0 horizon" comprises 239,770 data points for 7,776 binary mixtures, while an additional 19,937 experimental data points for 959 binary mixtures could only be predicted with mod.~UNIFAC 2.0 ("mod.~UNIFAC 2.0 only"). The boxes represent the interquartile ranges (IQR), and the whiskers extend to the last data points within 1.5 times the IQR from the box edges.}
    \label{modUNI20_fig:Performance_hE_bestModel}
\end{figure}

For evaluating the performance of mod.~UNIFAC 2.0 in predicting activity coefficients $\ln\gamma_i$ and excess enthalpies $h^\text{E}$, we use the mean absolute error (MAE) for each binary mixture and represent the results in box plots, as shown in~Figs.~\ref{modUNI20_fig:Performance_lngamma_bestModel} (for $\ln\gamma_i$) and \ref{modUNI20_fig:Performance_hE_bestModel} (for $h^\text{E}$). These plots also contain the corresponding results of mod.~UNIFAC 1.0, evaluated on the same basis, for comparison. The results shown in these figures were obtained with a mod.~UNIFAC 2.0 version trained on all available experimental data in our database. However, as detailed in the subsequent subsections, we have also performed two extrapolation tests by withholding parts of the data during the training to demonstrate and validate the predictive capacities of mod.~UNIFAC 2.0.

Although the exact training set for mod.~UNIFAC 1.0 has not been disclosed, it is reasonable to assume that the experimental data used in this work are similar to the data used for its parameterization, which supports a fair comparison in Figs.~\ref{modUNI20_fig:Performance_lngamma_bestModel} and \ref{modUNI20_fig:Performance_hE_bestModel}. Note that the comparison between mod.~UNIFAC 1.0 and mod.~UNIFAC 2.0 is carried out on the "mod.~UNIFAC 1.0 horizon", i.e., only those mixtures from our data set that can be modeled with the incomplete parameter set of mod.~UNIFAC 1.0. Since mod.~UNIFAC 2.0, with its completed parameter set, has a much larger scope, its performance is additionally evaluated on those mixtures that cannot be predicted with mod.~UNIFAC 1.0, labeled as the "mod.~UNIFAC 2.0 only" data set in Figs.~\ref{modUNI20_fig:Performance_lngamma_bestModel} and \ref{modUNI20_fig:Performance_hE_bestModel}.
The results in Fig.~\ref{modUNI20_fig:Performance_lngamma_bestModel} show an improved prediction accuracy for $\ln\gamma_i$ with  mod.~UNIFAC 2.0 compared to mod.~UNIFAC 1.0 for those mixtures that can be described with both models ("mod.~UNIFAC 1.0 horizon"). This is particularly evident concerning the mean of the MAE, which is nearly
halved with mod.~UNIFAC 2.0, demonstrating the ability of mod.~UNIFAC 2.0 to reduce very poorly predicted data points. Regarding the median of the MAE and the interquartile range, mod.~UNIFAC 2.0 also shows some improvements compared to mod.~UNIFAC 1.0.

These results indicate that using the holistic end-to-end training of mod.~UNIFAC 2.0 results in an improved set of pair-interaction parameters compared to the one obtained by the classical sequential fit carried out in the development of mod.~UNIFAC 1.0. However, the even more significant advantage of mod.~UNIFAC 2.0 is that its parameter set is complete, leading to a much broader applicability. By evaluating the results of mod.~UNIFAC 2.0 for those mixtures in our data set that cannot be modeled with mod.~UNIFAC 1.0 ("mod.~UNIFAC 2.0 only") in Fig.~\ref{modUNI20_fig:Performance_lngamma_bestModel}, we find a high prediction accuracy. It is similar to that of the results obtained with mod.~UNIFAC 1.0 for the mixtures to which this method can be applied.

Fig.~\ref{modUNI20_fig:Performance_hE_bestModel} shows the results for the prediction of $h^\text{E}$, where we see a similar picture as for the prediction of $\ln\gamma_i$: we find an improved performance of mod.~UNIFAC 2.0 on the "mod.~UNIFAC 1.0 horizon", and still high prediction accuracy on the "mod.~UNIFAC 2.0 only" data set, for which mod.~UNIFAC 1.0 cannot be applied.

\begin{figure}[H]
    \centering
    \includegraphics[width=\textwidth]{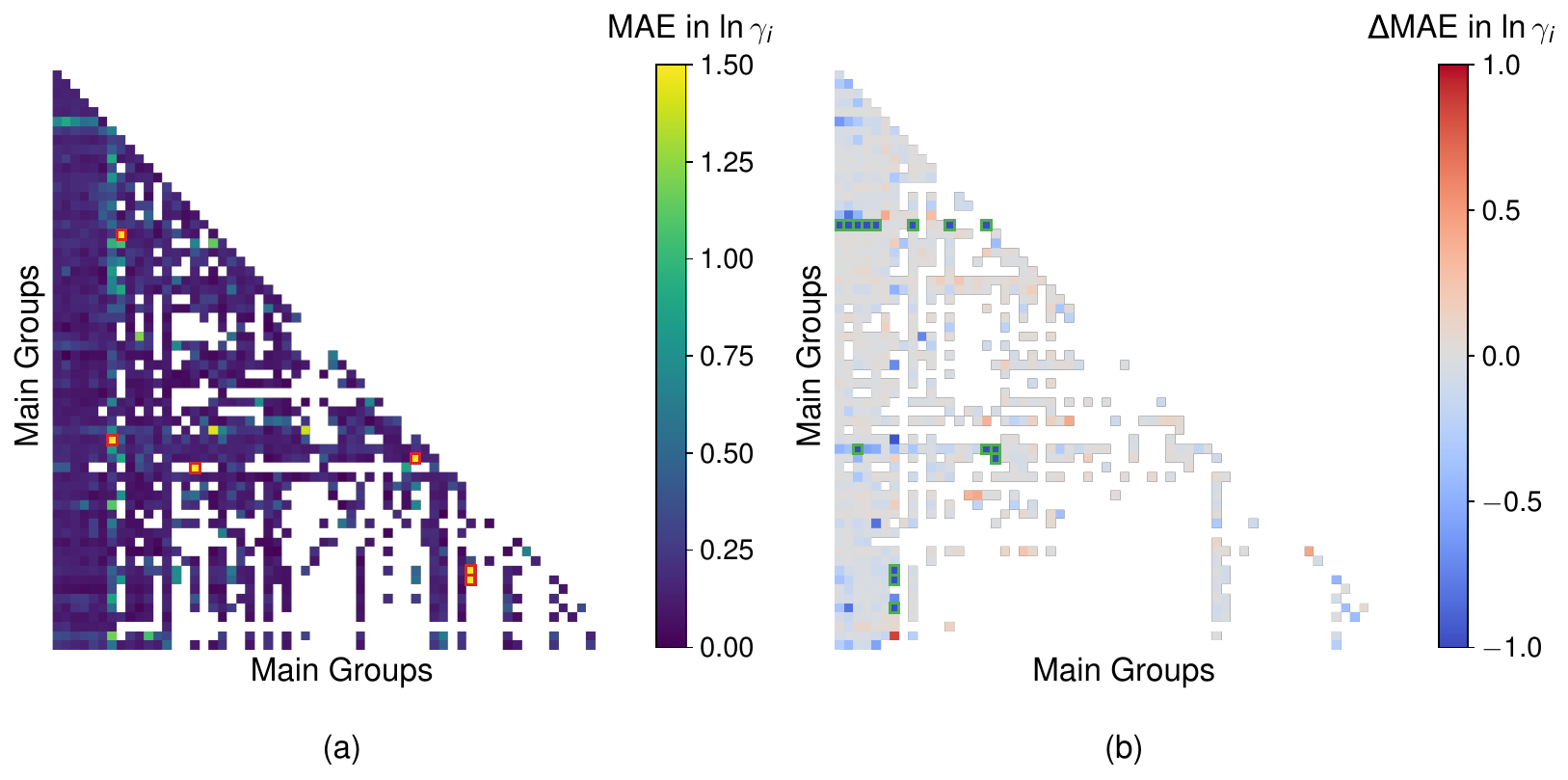}
    \caption{(a) Heatmap of the mean absolute error (MAE) of the predicted $\ln\gamma_i$ with mod.~UNIFAC 2.0 calculated for each pair of main groups by considering all data points for which that particular group combination is relevant. Group combinations with an MAE above 1.5 are highlighted by red frames. (b) Difference between the MAE in $\ln\gamma_i$ with mod.~UNIFAC 2.0 and the MAE of mod.~UNIFAC 1.0 on the "mod UNIFAC 1.0 horizon" ($\Delta\text{MAE} = \text{MAE}_{\text{mod.~UNIFAC 2.0}} - \text{MAE}_{\text{mod.~UNIFAC 1.0}}$) for each pair of main groups. Group combinations with a $\Delta$MAE below -1 are highlighted by green frames, indicating the most significant improvements with mod.~UNIFAC 2.0. Missing entries indicate that no data were available for the comparison of the given combination of groups.}
    \label{modUNI20_fig:Heatmaps_MAE_bestModel}
\end{figure}

Fig.~\ref{modUNI20_fig:Heatmaps_MAE_bestModel} provides a deeper insight into the overall performance of mod.~UNIFAC 2.0 by assigning an MAE for predicting $\ln\gamma_i$ to each pair of main groups, visualized as heatmaps. The shown MAEs are calculated by considering the predictions for all mixtures for which the respective group combination is relevant, with the number of mixtures and data points varying significantly among the pairs of main groups, as detailed in Fig.~S.1b of the Supporting Information. Panel (a) of Fig.~\ref{modUNI20_fig:Heatmaps_MAE_bestModel} shows the MAEs calculated as described above on our complete data set, while panel (b) visualizes improvements (or deteriorations) with mod.~UNIFAC 2.0 compared to mod.~UNIFAC 1.0 by showing the differences in the MAEs ($\Delta\text{MAE} = \text{MAE}_{\text{mod.~UNIFAC 2.0}} - \text{MAE}_{\text{mod.~UNIFAC 1.0}}$) on the "mod.~UNIFAC 1.0 horizon". Missing entries indicate that no data were available to compare the given combination of groups.
Fig.~\ref{modUNI20_fig:Heatmaps_MAE_bestModel}a highlights the overall strong performance of mod.~UNIFAC 2.0, with a small MAE for most group combinations. Note that the prediction for a particular mixture usually requires the consideration of multiple pair-interaction parameters. Hence, the MAEs in Fig.~\ref{modUNI20_fig:Heatmaps_MAE_bestModel}a, although assigned to specific group combinations, cannot be attributed to imperfections of the respective pair-interaction parameters alone, but are also affected by other pair-interaction parameters. However, despite this complexity, a clear trend can be observed. For instance, mixtures containing water (main group 7) apparently represent a particular challenge, likely because of the unique properties of water due to strong hydrogen bonding and polarity. While most group combinations yield an MAE below 0.14, which is a very good result, a few show high prediction errors. The three group combinations with an MAE greater than 2.0 are cases based on extremely small test sets, each consisting of a single binary mixture with ten or fewer data points. This suggests that the higher errors may also be due to limited data, and caution should be taken not to over-interpret these results.

Fig.~\ref{modUNI20_fig:Heatmaps_MAE_bestModel}b shows that mod.~UNIFAC 2.0 outperforms mod.~UNIFAC 1.0 for most group combinations. It significantly improves the results for 461 interaction parameters, with a mean $\Delta\text{MAE}$ of -0.31, whereas for the 267 combinations mod.~UNIFAC 1.0 yields better results; however, the deterioration is typically only minor with a mean $\Delta\text{MAE}$ of only 0.05. Notable improvements are observed for parameters involving main groups 7 ("H2O"), 18 ("PYRIDINE"), and 42 ("CY-CH2"), with ten group combinations showing extremely high MAE reductions with $\Delta\text{MAE} < -4$. In addition, parameters involving the most common group, main group 1 ("CH3"), also show significant improvements. For example, the mean MAE specific for the pair-interaction parameter between main group 1 ("CH3") and main group 7 ("H2O"), known to be poorly fitted in mod.~UNIFAC 1.0, is nearly halved, from 1.35 to 0.71.

\begin{figure}[H]
    \centering
    \includegraphics[width=\textwidth]{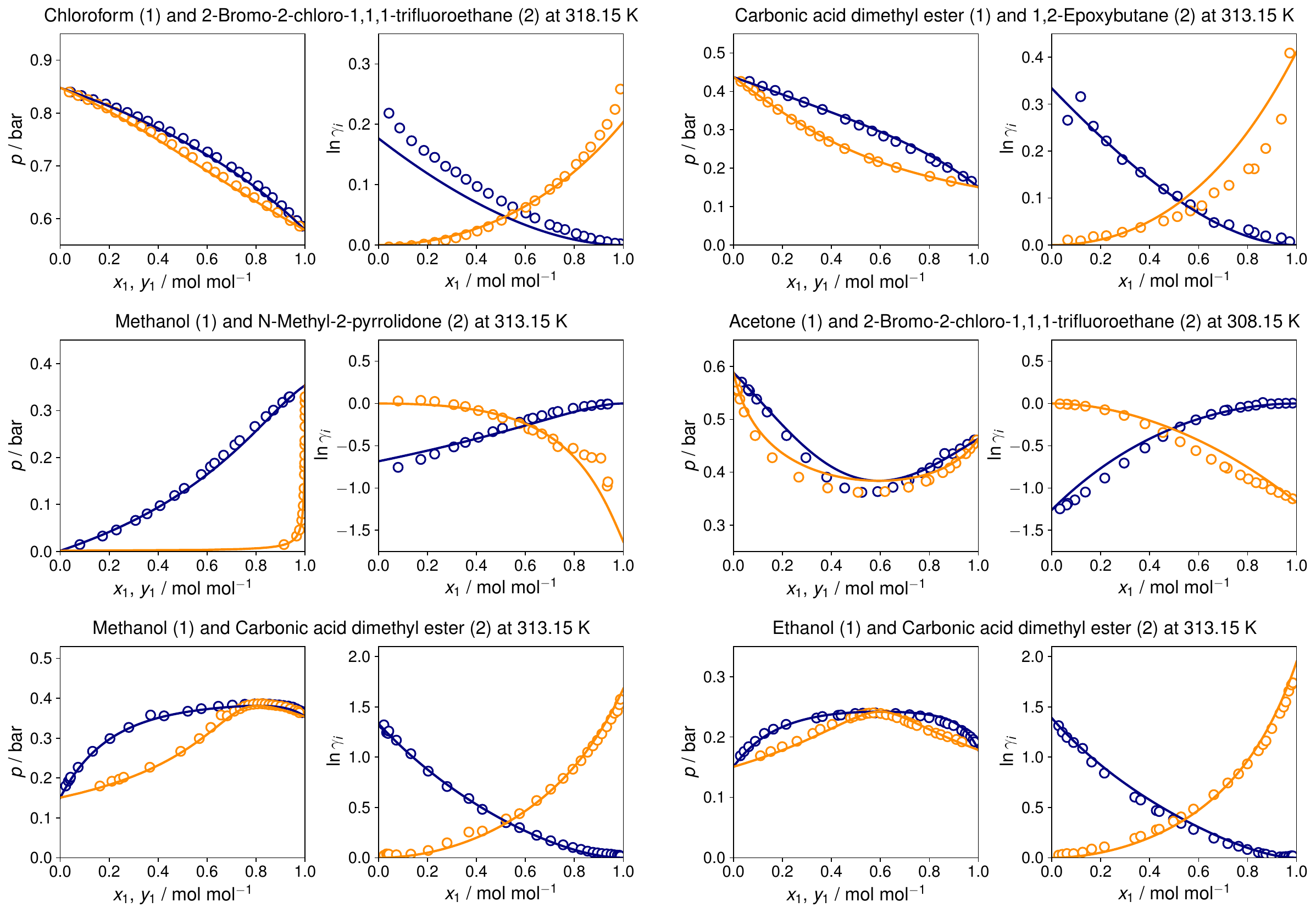}
    \caption{Prediction of $\ln\gamma_i$ and isothermal vapor–liquid phase diagrams for binary mixtures with mod.~UNIFAC 2.0 (lines) and comparison to experimental data from the DDB (symbols). Mod.~UNIFAC 1.0 is not applicable to the mixtures shown.}
    \label{modUNI20_fig:VLE_lngamma_binary}
\end{figure}

Fig.~\ref{modUNI20_fig:VLE_lngamma_binary} shows an example of the practical application of mod.~UNIFAC 2.0. It is used to predict vapor-liquid phase equilibria for binary mixtures, a critical task in chemical engineering. Six typical examples are shown, covering a range of phase behaviors from near-ideal mixtures to those with significant deviations, including low- and high-boiling azeotropes. All shown mixtures are part of the "mod.~UNIFAC 2.0 only" set, i.e., they cannot be modeled with mod.~UNIFAC 1.0. The predictions show an excellent agreement with the experimental data, underscoring the method's utility for modeling complex phase behavior and making it a valuable tool for a wide range of industrial processes, from distillation to solvent recovery.

\begin{figure}[H]
    \centering
    \includegraphics[width=\textwidth]{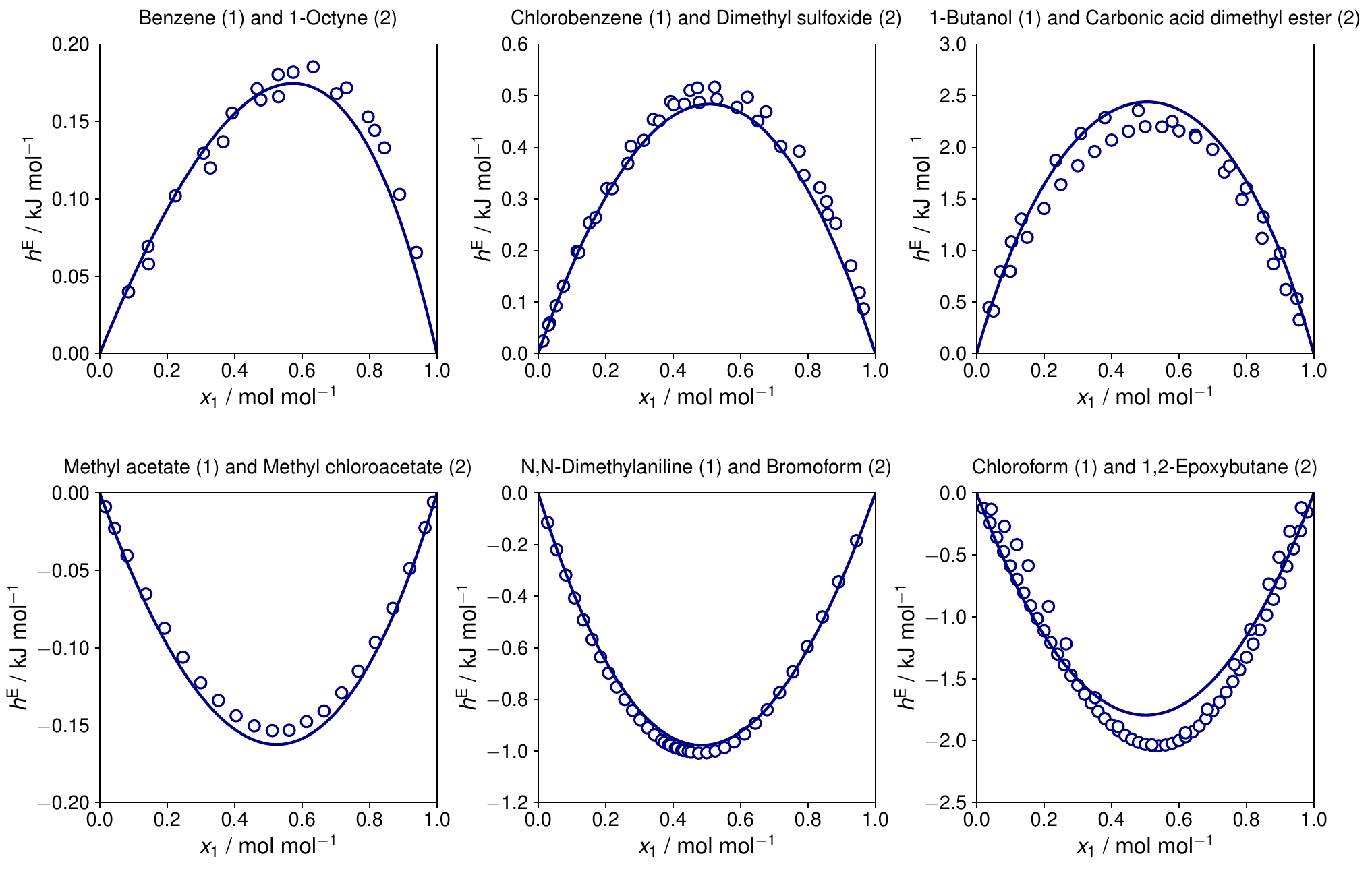}
    \caption{Prediction of excess enthalpies $h^\text{E}$ at 298.15~K for binary mixtures with mod.~UNIFAC 2.0 (lines) and comparison to experimental data from the DDB (symbols). Mod.~UNIFAC 1.0 is not applicable to the mixtures shown.}
    \label{modUNI20_fig:hE_binary}
\end{figure}

Fig.~\ref{modUNI20_fig:hE_binary} demonstrates the ability of mod.~UNIFAC 2.0 to predict excess enthalpies $h^\text{E}$ in binary mixtures. The figure presents six representative examples from the "mod.~UNIFAC 2.0 only" data set, i.e., mixtures for which mod.~UNIFAC 1.0 is not applicable, cf.~Fig.~\ref{modUNI20_fig:Performance_hE_bestModel}. The mixtures have been selected to highlight a variety of behaviors, ranging from nearly ideal to strongly non-ideal systems with both positive and negative deviations. The predicted excess enthalpy curves (solid lines) align closely with the experimental data (open circles), demonstrating the model's ability to accurately capture both the magnitude and the trend of $h^\text{E}$ across different mixtures.

Furthermore, since mod.~UNIFAC 2.0 is based on pairwise interactions between the structural groups occurring in the mixture, for any number of components, it allows for straightforward predictions of the properties of multi-component mixtures. In Fig.~S.3 in the Supporting Information, we show examples for modeling ternary mixtures with mod.~UNIFAC 2.0, which demonstrate its high predictive performance although being trained only on binary data.

\subsection{Extrapolation to Unseen Components}
\begin{figure}[H]
    \centering
    \includegraphics[width=0.8\textwidth]{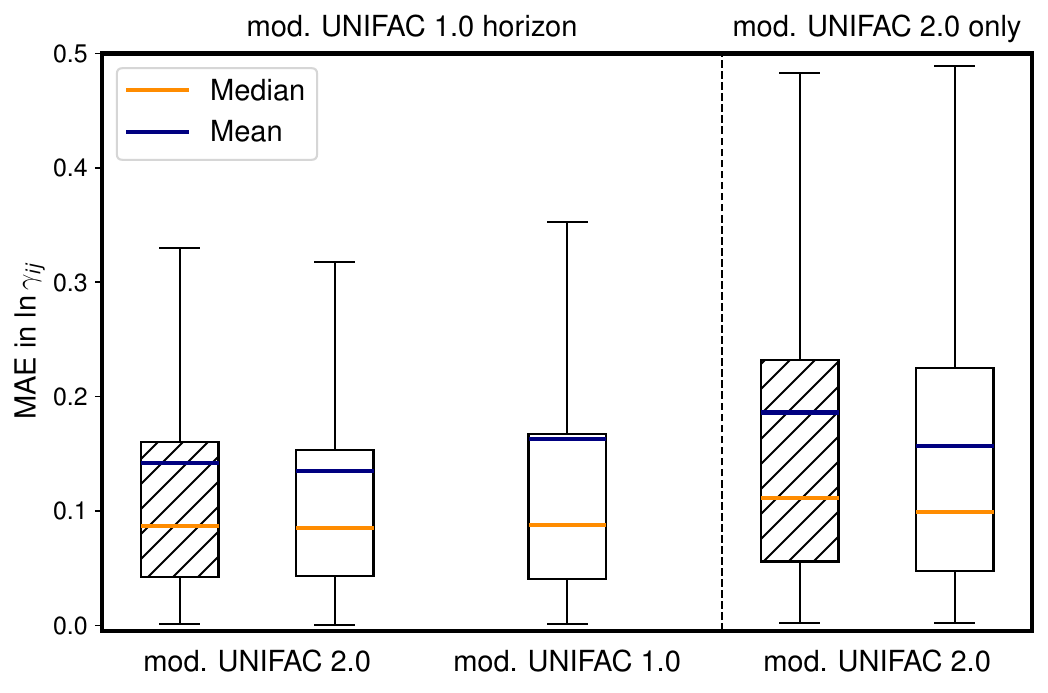}
    \caption{Mean absolute error (MAE) of the predicted $\ln\gamma_i$ of mixtures containing unseen components with mod.~UNIFAC 2.0 (shaded boxes). For comparison, the results of mod.~UNIFAC 2.0 trained on all experimental data and mod.~UNIFAC 1.0 are also shown (plain boxes). The "mod.~UNIFAC 1.0 horizon" comprises 19,015 data points for 1,254 binary mixtures, while an additional 1,897 experimental data points for 280 binary mixtures could only be predicted with mod.~UNIFAC 2.0 ("mod.~UNIFAC 2.0 only"). The boxes represent the interquartile ranges (IQR), and the whiskers extend to the last data points within 1.5 times the IQR from the box edges.}
    \label{modUNI20_fig:Performance_lngamma_unknownComp}
\end{figure}

To study the ability of mod.~UNIFAC 2.0 to extrapolate to mixtures involving components for which no mixture data were used in the training (termed "unseen components" in the following for simplicity), we have carried out a test in which 100 components were selected, and all data points containing any of these components were withheld from the training set and used only for testing the predictions. This exclusion resulted in a test set comprising 34,107 data points (20,912 for $\ln\gamma_i$ and 13,195 for $h^\text{E}$), covering 1,865 different binary mixtures. Fig.~\ref{modUNI20_fig:Performance_lngamma_unknownComp} shows the results for the prediction of $\ln\gamma_i$ for this test set, again represented as box plots of the mixture-specific MAE. Results from mod.~UNIFAC 1.0 are also shown for comparison.
Fig.~\ref{modUNI20_fig:Performance_lngamma_unknownComp} shows that the predictive accuracy of mod.~UNIFAC 2.0 for mixtures with components that were excluded from the training set (shaded boxes) is only narrowly lower than when the model is trained on the entire database (plain boxes). This consistency across both the "mod.~UNIFAC 1.0 horizon" and "mod.~UNIFAC 2.0 only" data sets underscores the robustness of the hybrid approach. Moreover, even on the test data, mod.~UNIFAC 2.0 outperforms mod.~UNIFAC 1.0 on the "mod.~UNIFAC 1.0 horizon", which is noteworthy given that mod.~UNIFAC 1.0 was likely trained on many of these test data points, as discussed earlier. On the "mod.~UNIFAC 2.0 only" data set, mod.~UNIFAC 2.0 shows slightly reduced predictive accuracy but still maintains strong performance, while mod.~UNIFAC 1.0 is not applicable. Overall, these results highlight the predictive power of mod.~UNIFAC 2.0. Similar trends were observed for the prediction of $h^\text{E}$, as shown in Fig.~S.4 in the Supporting Information.

\subsection{Extrapolation to Unseen Pair-Interaction Parameters}
Another, even more challenging, test to assess mod.~UNIFAC 2.0's predictive capacities is to test its ability to extrapolate to unseen pair interactions. For such a test, we have randomly selected 100 combinations of main groups, and have withheld all experimental data for mixtures for which the respective main group combinations are relevant from the training. For each of these 100 combinations, an individual test set was created from the withheld data, while all other available data were used to train mod.~UNIFAC 2.0. The number of data points and binary mixtures for the 100 test sets, as well as individual error scores, are given in Tables~S.1 (for $\ln\gamma_i$) and S.2 (for $h^\text{E}$) in the Supporting Information.

\begin{figure}[H]
    \centering
    \includegraphics[width=0.8\textwidth]{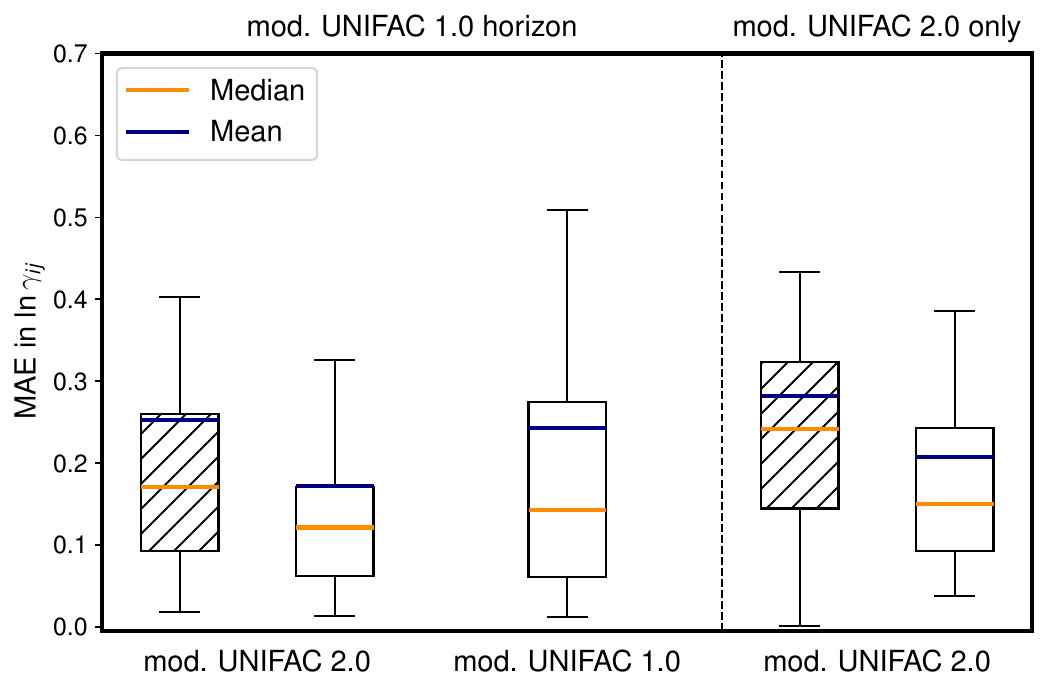}
    \caption{Mean absolute error (MAE) of the predicted $\ln\gamma_i$ with mod.~UNIFAC 2.0 for 100 test sets, where all data points for which a specific main group combination is relevant were withheld during training (shaded boxes); cf.~Table S.1 in the Supporting Information for numerical results. The results of mod.~UNIFAC 2.0 trained on all experimental data and mod.~UNIFAC 1.0 are shown for comparison (plain boxes). The boxes represent the interquartile ranges (IQR), and the whiskers extend to the last data points within 1.5 times the IQR from the box edges.}
    \label{modUNI20_fig:Performance_lngamma_loo}
\end{figure}

Fig.~\ref{modUNI20_fig:Performance_lngamma_loo} shows the results of predicting $\ln\gamma_i$ with mod.~UNIFAC 2.0 from this challenging test by summarizing the MAEs for the 100 test sets in a box plot. For comparison, the performance of mod.~UNIFAC 2.0 trained on all experimental data and the results of mod.~UNIFAC 1.0 (on the "mod.~UNIFAC 1.0 horizon") are included. Similar results were obtained for $h^\text{E}$ and are summarized in Fig.~S.6 in the Supporting Information.
The results on the "mod.~UNIFAC 1.0 horizon" demonstrate that mod.~UNIFAC 2.0, even when predicting truly unseen pair-interaction parameters, which could not directly be fitted to the training data, achieves a performance comparable to mod.~UNIFAC 1.0 with parameters that were likely fitted directly to the respective experimental data. Comparing mod.~UNIFAC 2.0's predictions for unseen pair interactions (shaded boxes) with those trained on the entire database (plain boxes) reveals a decrease in accuracy, as expected. However, the differences are modest, highlighting the robustness and reliability of mod.~UNIFAC 2.0 even in this extremely challenging test.

These tests emphasize the potential of mod.~UNIFAC 2.0 not only to broaden the applicability of this group-contribution method but also to improve its prediction accuracy significantly. Unlike mod.~UNIFAC 1.0, which is constrained by its limited parameter tables obtained from sequential fitting, mod.~UNIFAC 2.0 excels in both scope and performance, making it a robust tool for predicting activity coefficients across a wide range of mixtures. The superior accuracy demonstrated on the shared horizon confirms that mod.~UNIFAC 2.0 is not just a complementary option when mod.~UNIFAC 1.0 fails but a strong candidate to become the new standard.

Its ease of implementation sets mod.~UNIFAC 2.0 apart from other ML-based or hybrid models combining ML with physical modeling. Users can seamlessly adopt mod.~UNIFAC 2.0 by simply replacing the original parameter tables in their existing process simulators (or similar software), in which mod.~UNIFAC will most likely be implemented, with the completed parameter tables provided in the Supporting Information. This way, the tedious implementation of the ML model itself is eliminated, making mod.~UNIFAC 2.0 directly accessible for practical applications.

\clearpage
\section{Conclusions}
Mod.~UNIFAC~\cite{Constantinescu.2016} is currently the industrial standard for predicting activity coefficients and is implemented in basically all process simulation software packages. It is also widely used in academia and is the workhorse for calculating phase equilibria with liquid phases, such as vapor-liquid equilibria (VLE), liquid-liquid equilibria (LLE), and solid-liquid equilibria (SLE). Due to temperature-dependent parameters, it is more flexible than the original UNIFAC model~\cite{Wittig.2003} and often delivers better results. Furthermore, mod.~UNIFAC often gives better results than competing excess Gibbs energy models based on quantum-mechanical calculations of energetic contributions, such as COSMO-RS~\cite{Klamt.1995, Klamt.2000, Klamt.2005} and COSMO-SAC-dsp~\cite{Hsieh.2014}. 

However, mod.~UNIFAC has several important drawbacks. Firstly, the last published version stems from 2016 and has therefore been fitted only to data that were available up to then. Hence, the wealth of relevant data measured since then is omitted. More recent updates of mod.~UNIFAC are commercial and not publicly available. Secondly, and more importantly, as a group-contribution method, mod.~UNIFAC can only be applied to make predictions for a given mixture if all pair-interaction parameters (between all groups into which all components of the mixture are decomposed) are available. If only a single pair is missing, mod.~UNIFAC will not work. The latest public version of mod.~UNIFAC has 63 main groups, and, hence, 1953 pairs of groups -- but interaction parameters are only available for 756 of these pairs (39\%), which considerably limits the method's applicability. 

We have therefore developed mod.~UNIFAC 2.0, which overcomes these drawbacks: It was trained on data for activity coefficients and excess enthalpies published up to 2024 that were taken from the Dortmund Data Bank (DDB). All in all, more than 500,000 data points from 27,035 binary systems were used. The equations and the groups used in mod.~UNIFAC 2.0 are exactly the same as in the last published version, called mod.~UNIFAC 1.0 here, but the training differs drastically. While the parameters of mod.~UNIFAC 1.0 were determined in a sequential approach, without a chance to fill gaps for interactions for which no relevant data were available, mod.~UNIFAC 2.0 is trained using a matrix completion method (MCM) by which the entire interaction parameter matrix is filled simultaneously. Consequently, there are no gaps in the mod.~UNIFAC 2.0 parameter tables. This leads to an important extension of the applicability of the method. However, not only was the applicability extended, but the accuracy of the predictions was also improved. This was demonstrated in tests in which mod.~UNIFAC 2.0 was compared to mod.~UNIFAC 1.0: in different studies, data were deliberately excluded from the training and only used for the tests. Even in these tests, mod.~UNIFAC 2.0 performed consistently better than mod.~UNIFAC 1.0, even though they favor mod.~UNIFAC 1.0, as it must be assumed that relevant parts of the test set were used in its training. In-depth studies also reveal significant improvements for technically important classes of mixtures, such as mixtures containing water. 

As a method based on the physical concept of pair interactions, mod.~UNIFAC 2.0 can be used to predict thermodynamic properties not only for binary mixtures but also for multi-component mixtures. The new model can be seamlessly integrated into existing workflows, as users only need to update the parameter tables in existing implementations. Ultimately, the end-to-end training process of mod.~UNIFAC 2.0 allows for straightforward updates as new experimental data become available or for tailoring the model to specific industrial needs. Mod.~UNIFAC 2.0 demonstrates how combining machine learning with established physical models can significantly enhance the prediction of thermodynamic properties. Its expanded scope, improved accuracy, and ease of implementation represent a powerful and scalable solution for modern chemical engineering challenges. The complete parameter tables are freely provided in the Supporting Information as .csv files. We recommend using mod.~UNIFAC 2.0 as the default in all applications where, up to now, the default was mod.~UNIFAC 1.0.

\clearpage
\begin{acknowledgement}
We gratefully acknowledge financial support by Carl Zeiss Foundation in the project “Process Engineering 4.0”, as well as by Deutsche Forschungsgemeinschaft in the  Priority Program 2363, and in the Emmy Noether Project of FJ. 
\end{acknowledgement}


\clearpage
\providecommand{\latin}[1]{#1}
\makeatletter
\providecommand{\doi}
  {\begingroup\let\do\@makeother\dospecials
  \catcode`\{=1 \catcode`\}=2 \doi@aux}
\providecommand{\doi@aux}[1]{\endgroup\texttt{#1}}
\makeatother
\providecommand*\mcitethebibliography{\thebibliography}
\csname @ifundefined\endcsname{endmcitethebibliography}  {\let\endmcitethebibliography\endthebibliography}{}


\begin{mcitethebibliography}{30}
\providecommand*\natexlab[1]{#1}
\providecommand*\mciteSetBstSublistMode[1]{}
\providecommand*\mciteSetBstMaxWidthForm[2]{}
\providecommand*\mciteBstWouldAddEndPuncttrue
  {\def\EndOfBibitem{\unskip.}}
\providecommand*\mciteBstWouldAddEndPunctfalse
  {\let\EndOfBibitem\relax}
\providecommand*\mciteSetBstMidEndSepPunct[3]{}
\providecommand*\mciteSetBstSublistLabelBeginEnd[3]{}
\providecommand*\EndOfBibitem{}
\mciteSetBstSublistMode{f}
\mciteSetBstMaxWidthForm{subitem}{(\alph{mcitesubitemcount})}
\mciteSetBstSublistLabelBeginEnd
  {\mcitemaxwidthsubitemform\space}
  {\relax}
  {\relax}

\bibitem[Fredenslund \latin{et~al.}(1975)Fredenslund, Jones, and Prausnitz]{Fredenslund.1975}
Fredenslund,~A.; Jones,~R.~L.; Prausnitz,~J.~M. Group-contribution estimation of activity coefficients in nonideal liquid mixtures. \emph{AIChE Journal} \textbf{1975}, \emph{21}, 1086--1099\relax
\mciteBstWouldAddEndPuncttrue
\mciteSetBstMidEndSepPunct{\mcitedefaultmidpunct}
{\mcitedefaultendpunct}{\mcitedefaultseppunct}\relax
\EndOfBibitem
\bibitem[Magnussen \latin{et~al.}(1981)Magnussen, Rasmussen, and Fredenslund]{Magnussen.1981}
Magnussen,~T.; Rasmussen,~P.; Fredenslund,~A. UNIFAC parameter table for prediction of liquid-liquid equilibriums. \emph{Industrial {\&} Engineering Chemistry Process Design and Development} \textbf{1981}, \emph{20}, 331--339\relax
\mciteBstWouldAddEndPuncttrue
\mciteSetBstMidEndSepPunct{\mcitedefaultmidpunct}
{\mcitedefaultendpunct}{\mcitedefaultseppunct}\relax
\EndOfBibitem
\bibitem[Wittig \latin{et~al.}(2003)Wittig, Lohmann, and Gmehling]{Wittig.2003}
Wittig,~R.; Lohmann,~J.; Gmehling,~J. Vapor$-$Liquid Equilibria by UNIFAC Group Contribution. 6. Revision and Extension. \emph{Industrial {\&} Engineering Chemistry Research} \textbf{2003}, \emph{42}, 183--188\relax
\mciteBstWouldAddEndPuncttrue
\mciteSetBstMidEndSepPunct{\mcitedefaultmidpunct}
{\mcitedefaultendpunct}{\mcitedefaultseppunct}\relax
\EndOfBibitem
\bibitem[UNI(2023)]{UNIFAC_TUC.2023}
The UNIFAC Consortium. 2023; \url{http://www.unifac.org}\relax
\mciteBstWouldAddEndPuncttrue
\mciteSetBstMidEndSepPunct{\mcitedefaultmidpunct}
{\mcitedefaultendpunct}{\mcitedefaultseppunct}\relax
\EndOfBibitem
\bibitem[Constantinescu and Gmehling(2016)Constantinescu, and Gmehling]{Constantinescu.2016}
Constantinescu,~D.; Gmehling,~J. Further Development of Modified UNIFAC (Dortmund): Revision and Extension 6. \emph{Journal of Chemical {\&} Engineering Data} \textbf{2016}, \emph{61}, 2738--2748\relax
\mciteBstWouldAddEndPuncttrue
\mciteSetBstMidEndSepPunct{\mcitedefaultmidpunct}
{\mcitedefaultendpunct}{\mcitedefaultseppunct}\relax
\EndOfBibitem
\bibitem[{DDBST - Dortmund Data Bank Software {\&} Separation Technology GmbH}(2024)]{DDB.2024}
{DDBST - Dortmund Data Bank Software {\&} Separation Technology GmbH} Dortmund Data Bank. 2024; \url{www.ddbst.com}\relax
\mciteBstWouldAddEndPuncttrue
\mciteSetBstMidEndSepPunct{\mcitedefaultmidpunct}
{\mcitedefaultendpunct}{\mcitedefaultseppunct}\relax
\EndOfBibitem
\bibitem[{Chemstations, Inc.}(2024)]{ChemstationsInc..2024}
{Chemstations, Inc.} CHEMCAD V.8. 2024; \url{www.chemstations.com}\relax
\mciteBstWouldAddEndPuncttrue
\mciteSetBstMidEndSepPunct{\mcitedefaultmidpunct}
{\mcitedefaultendpunct}{\mcitedefaultseppunct}\relax
\EndOfBibitem
\bibitem[{Aspen Technology, Inc.}(2024)]{AspenTechnologyInc..2024}
{Aspen Technology, Inc.} Aspen Plus V14.5. 2024; \url{www.aspentech.com}\relax
\mciteBstWouldAddEndPuncttrue
\mciteSetBstMidEndSepPunct{\mcitedefaultmidpunct}
{\mcitedefaultendpunct}{\mcitedefaultseppunct}\relax
\EndOfBibitem
\bibitem[{A. Ramlatchan} \latin{et~al.}(2018){A. Ramlatchan}, {M. Yang}, {Q. Liu}, {M. Li}, {J. Wang}, and {Y. Li}]{A.Ramlatchan.2018}
{A. Ramlatchan}; {M. Yang}; {Q. Liu}; {M. Li}; {J. Wang}; {Y. Li} A survey of matrix completion methods for recommendation systems. \emph{Big Data Mining and Analytics} \textbf{2018}, \emph{1}, 308--323\relax
\mciteBstWouldAddEndPuncttrue
\mciteSetBstMidEndSepPunct{\mcitedefaultmidpunct}
{\mcitedefaultendpunct}{\mcitedefaultseppunct}\relax
\EndOfBibitem
\bibitem[Koren \latin{et~al.}(2009)Koren, Bell, and Volinsky]{Koren.2009}
Koren,~Y.; Bell,~R.; Volinsky,~C. Matrix Factorization Techniques for Recommender Systems. \emph{Computer} \textbf{2009}, \emph{42}, 30--37\relax
\mciteBstWouldAddEndPuncttrue
\mciteSetBstMidEndSepPunct{\mcitedefaultmidpunct}
{\mcitedefaultendpunct}{\mcitedefaultseppunct}\relax
\EndOfBibitem
\bibitem[Jirasek \latin{et~al.}(2020)Jirasek, Bamler, and Mandt]{Jirasek.2020}
Jirasek,~F.; Bamler,~R.; Mandt,~S. Hybridizing physical and data-driven prediction methods for physicochemical properties. \emph{Chemical Communications} \textbf{2020}, \emph{56}, 12407--12410\relax
\mciteBstWouldAddEndPuncttrue
\mciteSetBstMidEndSepPunct{\mcitedefaultmidpunct}
{\mcitedefaultendpunct}{\mcitedefaultseppunct}\relax
\EndOfBibitem
\bibitem[Jirasek \latin{et~al.}(2020)Jirasek, Alves, Damay, Vandermeulen, Bamler, Bortz, Mandt, Kloft, and Hasse]{Jirasek.2020b}
Jirasek,~F.; Alves,~R. A.~S.; Damay,~J.; Vandermeulen,~R.~A.; Bamler,~R.; Bortz,~M.; Mandt,~S.; Kloft,~M.; Hasse,~H. Machine Learning in Thermodynamics: Prediction of Activity Coefficients by Matrix Completion. \emph{The journal of physical chemistry letters} \textbf{2020}, \emph{11}, 981--985\relax
\mciteBstWouldAddEndPuncttrue
\mciteSetBstMidEndSepPunct{\mcitedefaultmidpunct}
{\mcitedefaultendpunct}{\mcitedefaultseppunct}\relax
\EndOfBibitem
\bibitem[Damay \latin{et~al.}(2021)Damay, Jirasek, Kloft, Bortz, and Hasse]{Damay.2021}
Damay,~J.; Jirasek,~F.; Kloft,~M.; Bortz,~M.; Hasse,~H. Predicting Activity Coefficients at Infinite Dilution for Varying Temperatures by Matrix Completion. \emph{Industrial {\&} Engineering Chemistry Research} \textbf{2021}, \emph{60}, 14564--14578\relax
\mciteBstWouldAddEndPuncttrue
\mciteSetBstMidEndSepPunct{\mcitedefaultmidpunct}
{\mcitedefaultendpunct}{\mcitedefaultseppunct}\relax
\EndOfBibitem
\bibitem[Hayer \latin{et~al.}(2022)Hayer, Jirasek, and Hasse]{Hayer.2022}
Hayer,~N.; Jirasek,~F.; Hasse,~H. Prediction of Henry's law constants by matrix completion. \emph{AIChE Journal} \textbf{2022}, \emph{68}, e17753\relax
\mciteBstWouldAddEndPuncttrue
\mciteSetBstMidEndSepPunct{\mcitedefaultmidpunct}
{\mcitedefaultendpunct}{\mcitedefaultseppunct}\relax
\EndOfBibitem
\bibitem[Gro{\ss}mann \latin{et~al.}(2022)Gro{\ss}mann, Bellaire, Hayer, Jirasek, and Hasse]{Gromann.2022}
Gro{\ss}mann,~O.; Bellaire,~D.; Hayer,~N.; Jirasek,~F.; Hasse,~H. Database for liquid phase diffusion coefficients at infinite dilution at 298 K and matrix completion methods for their prediction. \emph{Digital Discovery} \textbf{2022}, \emph{1}, 886--897\relax
\mciteBstWouldAddEndPuncttrue
\mciteSetBstMidEndSepPunct{\mcitedefaultmidpunct}
{\mcitedefaultendpunct}{\mcitedefaultseppunct}\relax
\EndOfBibitem
\bibitem[Jirasek \latin{et~al.}(2022)Jirasek, Bamler, Fellenz, Bortz, Kloft, Mandt, and Hasse]{Jirasek.2022}
Jirasek,~F.; Bamler,~R.; Fellenz,~S.; Bortz,~M.; Kloft,~M.; Mandt,~S.; Hasse,~H. Making thermodynamic models of mixtures predictive by machine learning: matrix completion of pair interactions. \emph{Chemical science} \textbf{2022}, \emph{13}, 4854--4862\relax
\mciteBstWouldAddEndPuncttrue
\mciteSetBstMidEndSepPunct{\mcitedefaultmidpunct}
{\mcitedefaultendpunct}{\mcitedefaultseppunct}\relax
\EndOfBibitem
\bibitem[Jirasek \latin{et~al.}(2023)Jirasek, Hayer, Abbas, Schmid, and Hasse]{Jirasek.2023}
Jirasek,~F.; Hayer,~N.; Abbas,~R.; Schmid,~B.; Hasse,~H. Prediction of parameters of group contribution models of mixtures by matrix completion. \emph{Physical chemistry chemical physics : PCCP} \textbf{2023}, \emph{25}, 1054--1062\relax
\mciteBstWouldAddEndPuncttrue
\mciteSetBstMidEndSepPunct{\mcitedefaultmidpunct}
{\mcitedefaultendpunct}{\mcitedefaultseppunct}\relax
\EndOfBibitem
\bibitem[Hayer \latin{et~al.}()Hayer, Wendel, Mandt, Hasse, and Jirasek]{Hayer.2024}
Hayer,~N.; Wendel,~T.; Mandt,~S.; Hasse,~H.; Jirasek,~F. Advancing Thermodynamic Group-Contribution Methods by Machine Learning: UNIFAC 2.0. \url{http://arxiv.org/pdf/2408.05220}\relax
\mciteBstWouldAddEndPuncttrue
\mciteSetBstMidEndSepPunct{\mcitedefaultmidpunct}
{\mcitedefaultendpunct}{\mcitedefaultseppunct}\relax
\EndOfBibitem
\bibitem[{DDBST - Dortmund Data Bank Software {\&} Separation Technology GmbH}(2023)]{DDB.2023}
{DDBST - Dortmund Data Bank Software {\&} Separation Technology GmbH} Dortmund Data Bank. 2023; \url{www.ddbst.com}\relax
\mciteBstWouldAddEndPuncttrue
\mciteSetBstMidEndSepPunct{\mcitedefaultmidpunct}
{\mcitedefaultendpunct}{\mcitedefaultseppunct}\relax
\EndOfBibitem
\bibitem[Gmehling \latin{et~al.}(2012)Gmehling, Kolbe, Kleiber, and Rarey]{Gmehling.2012}
Gmehling,~J.; Kolbe,~B.; Kleiber,~M.; Rarey,~J.~R. \emph{Chemical thermodynamics for process simulation}; Wiley-VCH-Verl.: Weinheim, 2012\relax
\mciteBstWouldAddEndPuncttrue
\mciteSetBstMidEndSepPunct{\mcitedefaultmidpunct}
{\mcitedefaultendpunct}{\mcitedefaultseppunct}\relax
\EndOfBibitem
\bibitem[Weidlich and Gmehling(1987)Weidlich, and Gmehling]{Weidlich.1987}
Weidlich,~U.; Gmehling,~J. A modified UNIFAC model. 1. Prediction of VLE, hE, and $\gamma^\infty$. \emph{Industrial {\&} Engineering Chemistry Research} \textbf{1987}, \emph{26}, 1372--1381\relax
\mciteBstWouldAddEndPuncttrue
\mciteSetBstMidEndSepPunct{\mcitedefaultmidpunct}
{\mcitedefaultendpunct}{\mcitedefaultseppunct}\relax
\EndOfBibitem
\bibitem[Gmehling \latin{et~al.}(1993)Gmehling, Li, and Schiller]{Gmehling.1993}
Gmehling,~J.; Li,~J.; Schiller,~M. A modified UNIFAC model. 2. Present parameter matrix and results for different thermodynamic properties. \emph{Industrial \& Engineering Chemistry Research} \textbf{1993}, \emph{32}, 178--193\relax
\mciteBstWouldAddEndPuncttrue
\mciteSetBstMidEndSepPunct{\mcitedefaultmidpunct}
{\mcitedefaultendpunct}{\mcitedefaultseppunct}\relax
\EndOfBibitem
\bibitem[Bingham \latin{et~al.}(2018)Bingham, Chen, Jankowiak, Obermeyer, Pradhan, Karaletsos, Singh, {Szerlip, Paul and Horsfall, Paul}, and Goodman]{Bingham.2018}
Bingham,~E.; Chen,~J.~P.; Jankowiak,~M.; Obermeyer,~F.; Pradhan,~N.; Karaletsos,~T.; Singh,~R.; {Szerlip, Paul and Horsfall, Paul}; Goodman,~N.~D. Pyro: Deep Universal Probabilistic Programming. \emph{Journal of Machine Learning Research} \textbf{2018}, \relax
\mciteBstWouldAddEndPunctfalse
\mciteSetBstMidEndSepPunct{\mcitedefaultmidpunct}
{}{\mcitedefaultseppunct}\relax
\EndOfBibitem
\bibitem[Blei \latin{et~al.}(2017)Blei, Kucukelbir, and McAuliffe]{Blei.2017}
Blei,~D.~M.; Kucukelbir,~A.; McAuliffe,~J.~D. Variational Inference: A Review for Statisticians. \emph{Journal of the American Statistical Association} \textbf{2017}, \emph{112}, 859--877\relax
\mciteBstWouldAddEndPuncttrue
\mciteSetBstMidEndSepPunct{\mcitedefaultmidpunct}
{\mcitedefaultendpunct}{\mcitedefaultseppunct}\relax
\EndOfBibitem
\bibitem[Kingma and Ba()Kingma, and Ba]{Kingma.22.12.2014}
Kingma,~D.~P.; Ba,~J. Adam: A Method for Stochastic Optimization. \url{http://arxiv.org/pdf/1412.6980.pdf}\relax
\mciteBstWouldAddEndPuncttrue
\mciteSetBstMidEndSepPunct{\mcitedefaultmidpunct}
{\mcitedefaultendpunct}{\mcitedefaultseppunct}\relax
\EndOfBibitem
\bibitem[Klamt(1995)]{Klamt.1995}
Klamt,~A. Conductor-like Screening Model for Real Solvents: A New Approach to the Quantitative Calculation of Solvation Phenomena. \emph{The Journal of Physical Chemistry} \textbf{1995}, \emph{99}, 2224--2235\relax
\mciteBstWouldAddEndPuncttrue
\mciteSetBstMidEndSepPunct{\mcitedefaultmidpunct}
{\mcitedefaultendpunct}{\mcitedefaultseppunct}\relax
\EndOfBibitem
\bibitem[Klamt and Eckert(2000)Klamt, and Eckert]{Klamt.2000}
Klamt,~A.; Eckert,~F. COSMO-RS: a novel and efficient method for the a priori prediction of thermophysical data of liquids. \emph{Fluid Phase Equilibria} \textbf{2000}, \emph{172}, 43--72\relax
\mciteBstWouldAddEndPuncttrue
\mciteSetBstMidEndSepPunct{\mcitedefaultmidpunct}
{\mcitedefaultendpunct}{\mcitedefaultseppunct}\relax
\EndOfBibitem
\bibitem[Klamt(2005)]{Klamt.2005}
Klamt,~A. \emph{COSMO-RS: From quantum chemistry to fluid phase thermodynamics and drug design}, 1st ed.; Elsevier: Amsterdam, 2005\relax
\mciteBstWouldAddEndPuncttrue
\mciteSetBstMidEndSepPunct{\mcitedefaultmidpunct}
{\mcitedefaultendpunct}{\mcitedefaultseppunct}\relax
\EndOfBibitem
\bibitem[Hsieh \latin{et~al.}(2014)Hsieh, Lin, and Vrabec]{Hsieh.2014}
Hsieh,~C.-M.; Lin,~S.-T.; Vrabec,~J. Considering the dispersive interactions in the COSMO-SAC model for more accurate predictions of fluid phase behavior. \emph{Fluid Phase Equilibria} \textbf{2014}, \emph{367}, 109--116\relax
\mciteBstWouldAddEndPuncttrue
\mciteSetBstMidEndSepPunct{\mcitedefaultmidpunct}
{\mcitedefaultendpunct}{\mcitedefaultseppunct}\relax
\EndOfBibitem
\end{mcitethebibliography}
\end{document}


%
%
%
%
%

\section{Modified UNIFAC Parameterization}
\begin{figure}[H]
    \centering
    \includegraphics[width=\textwidth]{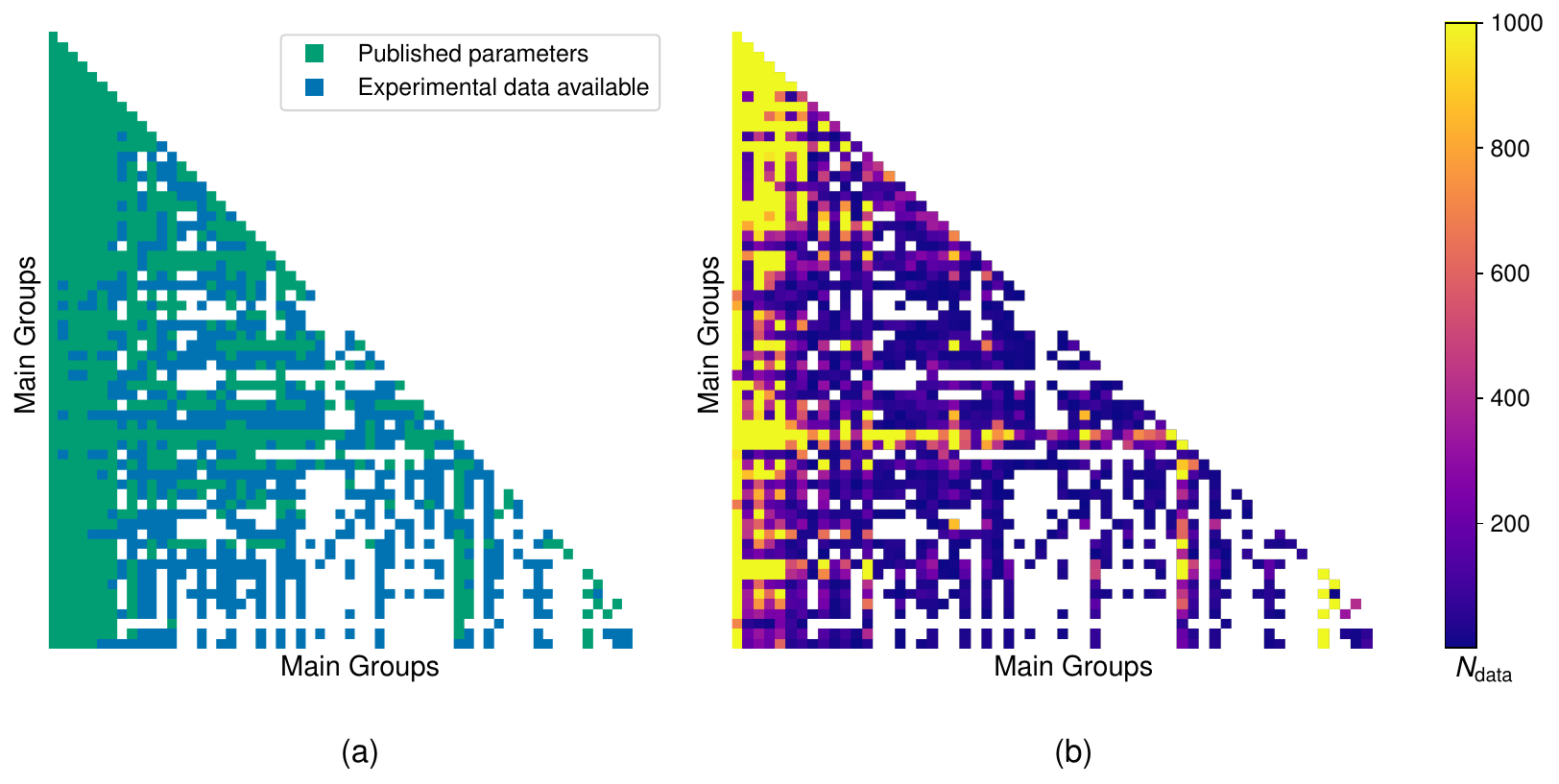}
    \caption{(a) Representation of the published mod.~UNIFAC 1.0 pair-interaction parameters~\cite{Constantinescu.2016} (green) and the ones that could additionally be fitted using the experimental data from the DDB~\cite{DDB.2024} (blue). (b) Heatmap of the number of experimental data points ($\ln\gamma_i$ and $h^\text{E}$) from the DDB requiring specific main group combinations.}
    \label{modUNI20_fig:Heatmaps_DataBase}
\end{figure}

Fig.~\ref{modUNI20_fig:Heatmaps_DataBase}a provides a visualization of the pair-interaction parameters included in the standard mod.~UNIFAC 1.0 model~\cite{Constantinescu.2016}, as well as those that can be additionally fitted using the experimental database discussed in the "Data" section. Additionally, Fig.~\ref{modUNI20_fig:Heatmaps_DataBase}b illustrates the distribution of the experimental data points considered here associated with each main group combination, highlighting the extent to which the respective pair-interaction parameters are supported by the available data.
The heatmaps reveal a strong disparity in data coverage. For example, while 216 group combinations (11\% of the matrix) are associated with more than 1,000 data points, 248 group combinations (13\%) are only relevant in 20 or fewer data points. Even worse, 594 parameters (30\%) do not appear in the available experimental data and can not be directly fitted. This pronounced heterogeneity underscores the challenges of parameter fitting and emphasizes the importance of models like mod.~UNIFAC 2.0 that efficiently use the available experimental data to fill the gaps.

\begin{figure}[H]
    \centering
    \includegraphics[width=0.5\textwidth]{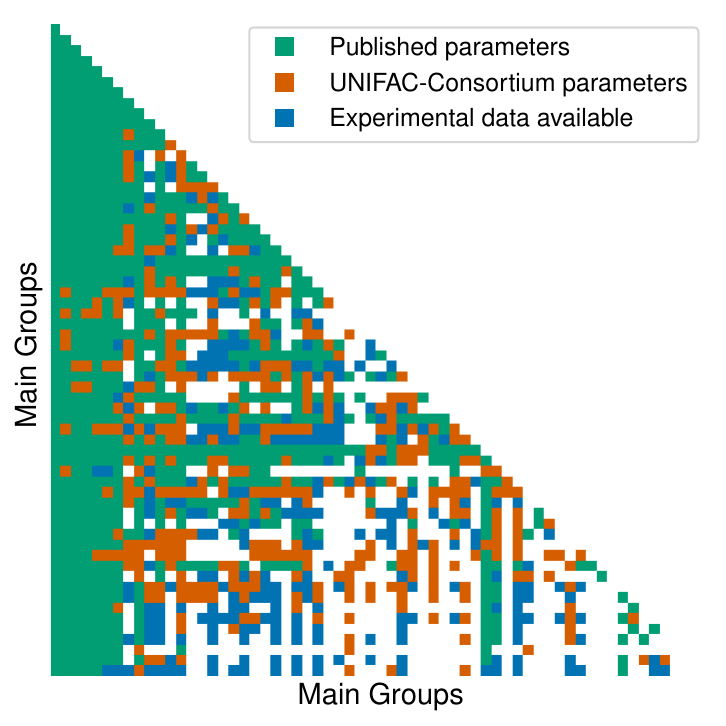}
    \caption{Matrix of existing pair-interaction parameters of the public mod.~UNIFAC 1.0 model~\cite{Constantinescu.2016} (green), supplemented by those of the commercial UNIFAC-Consortium version~\cite{UNIFAC_TUC.2023} (orange). Furthermore, group combinations are marked for which data are available, but no parameters have yet been fitted (blue).}
    \label{modUNI20_fig:Heatmap_Consortium}
\end{figure}

Fig.~\ref{modUNI20_fig:Heatmap_Consortium} extends Fig.\ref{modUNI20_fig:Heatmaps_DataBase}a by incorporating the interaction parameters available in the commercial UNIFAC-Consortium model~\cite{UNIFAC_TUC.2023}. It is important to note that the UNIFAC-Consortium version includes more main groups than the 63 considered in the public version, which have been omitted for consistency here.
While the UNIFAC-Consortium model provides a substantially broader scope than the public mod.~UNIFAC 1.0, the heatmap in Fig.~\ref{modUNI20_fig:Heatmap_Consortium} still highlights significant gaps in the interaction parameter matrix. These gaps are primarily due to the limited availability of experimental training data, emphasizing the critical need for extrapolative methods such as mod.~UNIFAC 2.0. Unlike traditional models, mod.~UNIFAC 2.0 can predict these missing parameters, thus bridging the gaps in the interaction parameter space. However, since the parameter tables for the UNIFAC-Consortium model are proprietary, a direct evaluation or comparison of its predictive accuracy could not be performed in this study.

\section{Extrapolation to Multi-Component Mixtures}
\begin{figure}[H]
    \centering
    \includegraphics[width=\textwidth]{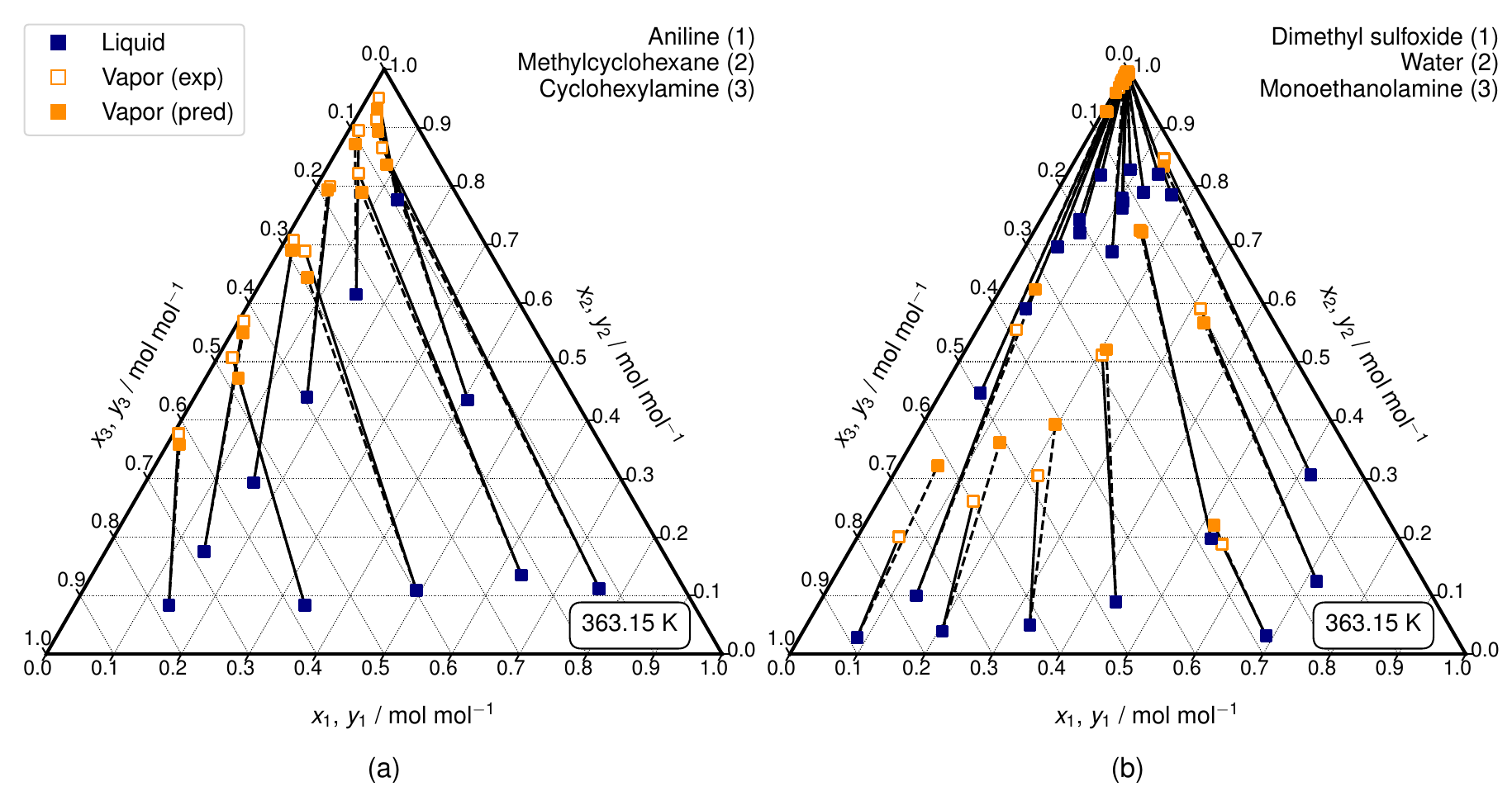}
    \caption{Prediction of isothermal vapor-liquid phase diagrams for ternary mixtures with mod.~UNIFAC 2.0 (pred) and comparison to experimental data (exp) from the DDB. The temperature and the composition of the liquid phase were specified, and the composition of the corresponding vapor phase in equilibrium was predicted. Solid lines are experimental conodes, dashed lines are predicted conodes. Mod.~UNIFAC 1.0 is not applicable to the mixtures shown.}
    \label{modUNI20_fig:VLE_ternary}
\end{figure}

Despite the absence of multi-component mixture data during the training of mod.~UNIFAC 2.0, the physical principles of its framework allow it to make reliable predictions for such systems. To illustrate this capability, Fig.~\ref{modUNI20_fig:VLE_ternary} shows isothermal vapor-liquid phase diagrams for two ternary mixtures from the "mod.~UNIFAC 2.0 only" data set, i.e., mixtures for which mod.~UNIFAC 1.0 cannot be applied due to missing pair-interaction parameters.

In these examples, the temperature and liquid-phase composition (shown as blue symbols in Fig.~\ref{modUNI20_fig:VLE_ternary}) were used as inputs to mod.~UNIFAC 2.0. The model predicted the activity coefficients from which the corresponding vapor-phase composition at equilibrium (filled orange symbols) could be calculated using the extended Raoult's law, cf.~Eq.~(8) in the manuscript. For comparison, the experimental vapor-phase compositions are also shown (open orange symbols). The predictions are in excellent agreement with the experimental data, demonstrating the model's suitability for describing multi-component mixtures.

\clearpage
\section{Prediction of Excess Enthalpies for Unseen Components}
\begin{figure}[H]
    \centering
    \includegraphics[width=0.8\textwidth]{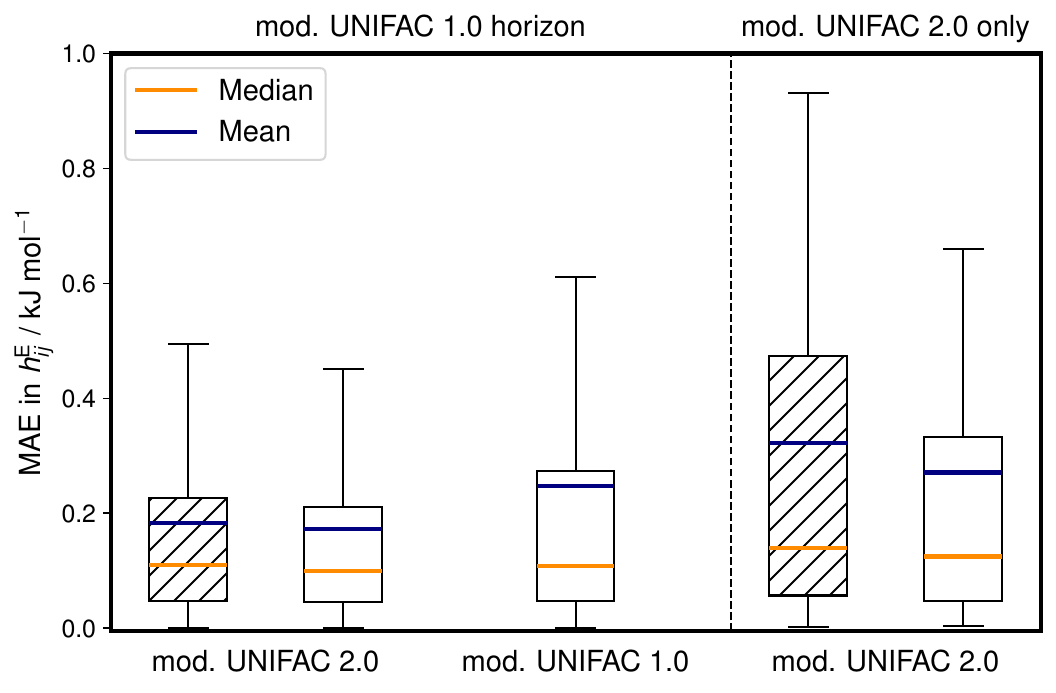}
    \caption{Mean absolute error (MAE) of the predicted $h^\text{E}$ of mixtures containing unseen components with mod.~UNIFAC 2.0 (shaded boxes). For comparison, the results of mod.~UNIFAC 2.0 trained on all experimental data and mod.~UNIFAC 1.0 are also shown (plain boxes). The "mod.~UNIFAC 1.0 horizon" comprises 11,906 data points for 473 binary mixtures, while an additional 1,289 experimental data points for 71 binary mixtures could only be predicted with mod.~UNIFAC 2.0 ("mod.~UNIFAC 2.0 only"). The boxes represent the interquartile ranges (IQR), and the whiskers extend to the last data points within 1.5 times the IQR from the box edges.}
    \label{modUNI20_fig:Performance_hE_unknownComp}
\end{figure}

The extrapolation capability of mod.~UNIFAC 2.0 for mixtures containing unseen components was evaluated by randomly selecting 100 components and training the model on all available data ($\ln\gamma_i$ and $h^\text{E}$) for those mixtures where these components do not occur. The retained mixtures served as the test set. Fig.~\ref{modUNI20_fig:Performance_hE_unknownComp} presents results for predicting $h^\text{E}$ on this test set in box plots, analogous to Fig.~7 in the manuscript, which focuses on $\ln\gamma_i$.
On the "mod.~UNIFAC 1.0 horizon", mod.~UNIFAC 2.0 not only outperforms mod.~UNIFAC 1.0 but also achieves comparable prediction accuracies whether trained on all experimental data or tasked with true extrapolation, demonstrating its robust extrapolation capabilities. On the "mod.~UNIFAC 2.0 only" set, slightly higher MAE values are observed. However, the method still provides reasonable predictions for most mixtures, as evidenced by the low median, underscoring its applicability in scenarios where mod.~UNIFAC 1.0 cannot be applied.

\section{Extrapolation to Unseen Pair-Interaction Parameters}
\begin{figure}[H]
    \centering
    \includegraphics[width=0.5\textwidth]{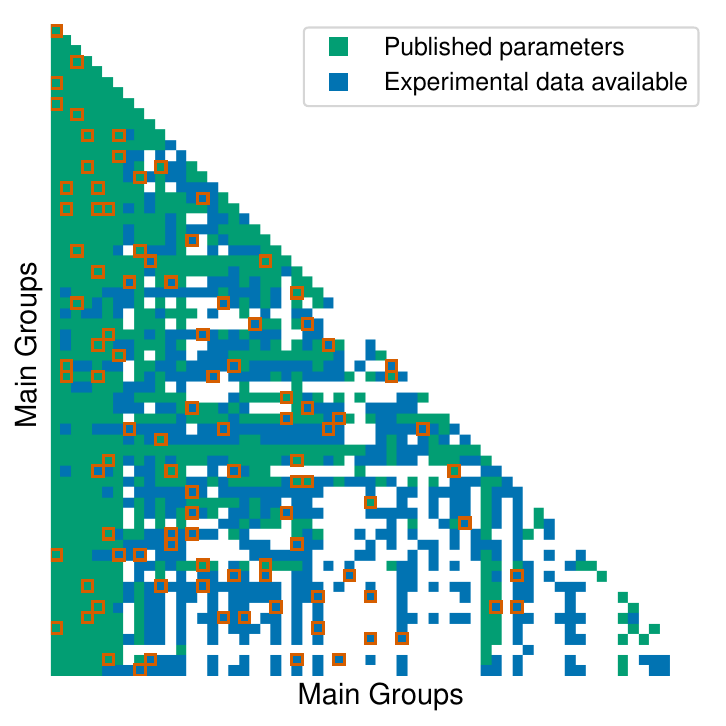}
    \caption{Matrix of available pair-interaction parameters of the mod.~UNIFAC 1.0 model~\cite{Constantinescu.2016} (green) alongside additional group combinations for which experimental data are available~\cite{DDB.2024} (blue). Group combinations that have been selected for the extrapolation study in this work are highlighted by orange frames, cf.~Tables~\ref{modUNI20_tab:loo_sets} and \ref{modUNI20_tab:loo_sets_hE}.}
    \label{modUNI20_fig:Heatmap_loo}
\end{figure}

Fig.~\ref{modUNI20_fig:Heatmap_loo} shows the 100 main group combinations randomly selected for the second extrapolation study, cf.~Fig.~8 in the manuscript. For each combination, all $\ln\gamma_i$ and $h^\text{E}$ data associated with the corresponding interaction parameters were removed from the training set and used exclusively for testing. The frequency of these group combinations varies in the experimental database, resulting in test sets of different sizes and compositions. A comprehensive summary of all 100 test sets, including the number of data points and mixtures, is provided in Tables~\ref{modUNI20_tab:loo_sets} (for $\ln\gamma_i$) and \ref{modUNI20_tab:loo_sets_hE} (for $h^\text{E}$). These tables also include MAEs for each individual test set.
In this extrapolation study, 53 of the selected group combinations were not parameterized in mod.~UNIFAC 1.0. Moreover, even when these parameters are available, they do not guarantee that all binary mixtures within the test sets can be predicted, as additional interaction parameters might also be required. To address this distinction, Tables~\ref{modUNI20_tab:loo_sets} and \ref{modUNI20_tab:loo_sets_hE} categorize the data into two groups: mixtures that can be predicted by both mod.~UNIFAC 1.0 and mod.~UNIFAC 2.0 (referred to as the "mod.~UNIFAC 1.0 horizon") and those that can only be predicted by mod.~UNIFAC 2.0 ("UNIFAC 2.0 only"). Since certain group combinations have data for either $\ln\gamma_i$ or $h^\text{E}$ (but not both), they appear in only one of the following tables.
\begin{longtable}{l r r r r r r r}
    \caption{Test sets for $\ln\gamma_i$ evaluated for predicting interaction parameters. Each set is categorized into two groups: "mod.~UNIFAC 1.0 horizon" and "mod.~UNIFAC 2.0 only". The main group identifiers ($m-n$) are identical to mod.~UNIFAC 1.0~\cite{Constantinescu.2016}. The table lists the number of data points ($N_\text{data}$) and binary mixtures ($N_\text{mix}$) for each set. It also includes the mixture-wise mean absolute errors, MAE$^\text{1.0}_\text{mix}$ and MAE$^\text{2.0}_\text{mix}$, for both mod.~UNIFAC methods.}\\
    \toprule
    \multirow{2}{*}{$m-n$} & \multicolumn{4}{c}{mod.~UNIFAC 1.0 horizon} & \multicolumn{3}{c}{mod.~UNIFAC 2.0 only} \\ \cline{2-8}
    & $N_\text{data}$ & $N_\text{mix}$ & MAE$^\text{2.0}_\text{mix}$ & MAE$^\text{1.0}_\text{mix}$ & $N_\text{data}$ & $N_\text{mix}$ & MAE$^\text{2.0}_\text{mix}$ \\
    \midrule
    \endfirsthead

    \caption*{Table \ref{modUNI20_tab:loo_sets} continued.}\\
    \toprule
    \multirow{2}{*}{$m-n$} & \multicolumn{4}{c}{mod.~UNIFAC 1.0 horizon} & \multicolumn{3}{c}{mod.~UNIFAC 2.0 only} \\ \cline{2-8}
    & $N_\text{data}$ & $N_\text{mix}$ & MAE$^\text{2.0}_\text{mix}$ & MAE$^\text{1.0}_\text{mix}$ & $N_\text{data}$ & $N_\text{mix}$ & MAE$^\text{2.0}_\text{mix}$ \\
    \midrule
    \endhead
    
1-2 & 21164 & 3021 & 0.18 & 0.22 & 1617 & 535 & 0.21\\
1-7 & 19331 & 680 & 1.18 & 1.35 & 514 & 58 & 1.27\\
1-9 & 19557 & 1557 & 0.17 & 0.16 & 976 & 153 & 0.38\\
1-52 & 672 & 26 & 0.06 & 0.08 & 1447 & 58 & 0.18\\
1-59 & 1860 & 268 & 0.24 & 0.47 & 1115 & 224 & 0.26\\
2-17 & 92 & 20 & 0.09 & 0.91 & 15 & 13 & 0.25\\
2-19 & 741 & 275 & 0.19 & 0.14 & 131 & 101 & 0.11\\
2-34 & 206 & 21 & 0.08 & 0.05 & 85 & 25 & 0.25\\
2-35 & 37 & 15 & 0.16 & 0.16 & 19 & 10 & 0.41\\
3-5 & 10550 & 795 & 0.23 & 0.51 & 367 & 63 & 0.22\\
3-10 & 672 & 69 & 0.19 & 0.2 & 192 & 71 & 0.28\\
3-23 & 45 & 4 & 0.24 & 0.02 & 26 & 9 & 0.14\\
3-28 & 82 & 7 & 0.05 & 0.05 & 3 & 3 & 0.27\\
4-12 & 102 & 7 & 0.13 & 0.1 &  &  & \\
4-15 & 136 & 15 & 0.1 & 0.13 & 70 & 4 & 0.19\\
4-55 & 1971 & 284 & 0.2 & 0.18 & 632 & 145 & 0.18\\
4-58 & 450 & 70 & 0.37 & 0.34 & 109 & 37 & 0.18\\
5-17 & 97 & 6 & 0.32 & 0.13 &  &  & \\
5-19 & 994 & 70 & 0.17 & 0.17 & 29 & 8 & 0.14\\
5-25 & 818 & 29 & 0.11 & 0.11 & 25 & 5 & 0.08\\
5-32 & 50 & 14 & 0.06 & 0.06 &  &  & \\
5-35 & 200 & 8 & 0.29 & 0.21 & 48 & 5 & 0.28\\
5-57 & 143 & 35 & 0.05 & 0.07 & 42 & 7 & 0.08\\
6-19 & 269 & 8 & 0.17 & 0.12 & 7 & 3 & 0.11\\
6-31 & 69 & 1 & 0.22 & 0.17 &  &  & \\
6-43 & 372 & 7 & 0.13 & 0.14 &  &  & \\
6-50 & 88 & 1 & 0.09 & 0.03 &  &  & \\
6-62 & 56 & 10 & 0.35 & 0.46 & 13 & 3 & 0.31\\
7-12 & 136 & 14 & 0.56 & 0.31 &  &  & \\
7-14 & 1313 & 34 & 0.85 & 0.46 &  &  & \\
7-33 & 166 & 27 & 1.06 & 0.98 & 6 & 1 & 0.62\\
7-52 &  &  &  &  & 21 & 2 & 0.32\\
8-26 &  &  &  &  & 4 & 4 & 0.42\\
8-40 &  &  &  &  & 26 & 18 & 0.31\\
9-16 & 102 & 28 & 0.09 & 0.19 & 11 & 5 & 0.11\\
9-23 & 48 & 2 & 0.06 & 0.06 &  &  & \\
9-52 &  &  &  &  & 260 & 8 & 0.1\\
9-63 &  &  &  &  & 83 & 13 & 0.17\\
10-24 &  &  &  &  & 11 & 1 & 0.41\\
10-62 &  &  &  &  & 30 & 6 & 0.2\\
11-15 & 377 & 1 & 0.06 & 0.03 &  &  & \\
11-41 & 346 & 105 & 0.21 & 0.27 & 53 & 21 & 0.24\\
11-55 &  &  &  &  & 517 & 95 & 0.38\\
12-44 & 135 & 5 & 0.26 & 0.23 &  &  & \\
12-51 &  &  &  &  & 11 & 1 & 0.38\\
14-22 &  &  &  &  & 11 & 3 & 0.17\\
14-50 &  &  &  &  & 6 & 1 & 0.39\\
15-18 &  &  &  &  & 14 & 1 & 0.12\\
15-31 &  &  &  &  & 75 & 2 & 0.26\\
16-35 &  &  &  &  & 1 & 1 & 0.37\\
17-28 &  &  &  &  & 1 & 1 & 0.19\\
17-40 &  &  &  &  & 42 & 26 & 0.35\\
18-34 &  &  &  &  & 4 & 4 & 0.27\\
18-54 &  &  &  &  & 31 & 1 & 0.26\\
19-58 &  &  &  &  & 72 & 13 & 0.19\\
20-30 &  &  &  &  & 40 & 3 & 0.68\\
21-24 & 228 & 11 & 0.14 & 0.1 & 22 & 1 & 0.14\\
21-53 & 9 & 2 & 0.4 & 0.03 &  &  & \\
21-54 &  &  &  &  & 3 & 3 & 0.29\\
22-57 &  &  &  &  & 24 & 7 & 0.15\\
23-37 & 36 & 3 & 0.09 & 0.05 & 4 & 2 & 1.8\\
23-48 &  &  &  &  & 13 & 11 & 0.32\\
24-27 & 18 & 1 & 0.6 & 0.49 &  &  & \\
24-43 & 192 & 2 & 0.02 & 0.03 &  &  & \\
24-45 & 127 & 1 & 0.14 & 0.01 &  &  & \\
24-51 &  &  &  &  & 42 & 2 & 0.05\\
24-62 &  &  &  &  & 30 & 6 & 0.3\\
25-30 &  &  &  &  & 32 & 1 & 0.05\\
25-38 &  &  &  &  & 40 & 8 & 0.1\\
25-45 & 19 & 1 & 0.12 & 0.08 & 3 & 1 & 0.76\\
26-56 &  &  &  &  & 184 & 44 & 0.18\\
26-59 &  &  &  &  & 34 & 7 & 0.11\\
27-32 &  &  &  &  & 3 & 2 & 0.27\\
27-40 &  &  &  &  & 182 & 101 & 0.21\\
28-39 &  &  &  &  & 1 & 1 & 0.25\\
28-62 &  &  &  &  & 15 & 3 & 0.43\\
29-54 &  &  &  &  & 24 & 2 & 0.08\\
31-47 & 10 & 1 & 0.08 & 0.02 &  &  & \\
33-34 &  &  &  &  & 1 & 1 & 0.15\\
33-35 &  &  &  &  & 1 & 1 & 0\\
34-60 &  &  &  &  & 78 & 13 & 0.33\\
36-40 &  &  &  &  & 3 & 1 & 0.19\\
39-44 & 128 & 2 & 0.82 & 0.84 &  &  & \\
43-57 &  &  &  &  & 50 & 11 & 0.11\\
45-54 &  &  &  &  & 17 & 2 & 0.4\\
45-57 &  &  &  &  & 24 & 7 & 0.11\\
\bottomrule
\label{modUNI20_tab:loo_sets}
\end{longtable}

\begin{longtable}{l r r r r r r r}
    \caption{Test sets for $h^\text{E}$ evaluated for predicting interaction parameters. Each set is categorized into two groups: "mod.~UNIFAC 1.0 horizon" and "mod.~UNIFAC 2.0 only". The main group identifiers ($m-n$) are identical to mod.~UNIFAC 1.0~\cite{Constantinescu.2016}. The table lists the number of data points ($N_\text{data}$) and binary mixtures ($N_\text{mix}$) for each set. It also includes the mixture-wise mean absolute errors, MAE$^\text{1.0}_\text{mix}$ and MAE$^\text{2.0}_\text{mix}$, for both mod.~UNIFAC methods.}\\
	\toprule
    \multirow{2}{*}{$m-n$} & \multicolumn{4}{c}{mod.~UNIFAC 1.0 horizon} & \multicolumn{3}{c}{mod.~UNIFAC 2.0 only} \\ \cline{2-8}
    & $N_\text{data}$ & $N_\text{mix}$ & MAE$^\text{2.0}_\text{mix}$ & MAE$^\text{1.0}_\text{mix}$ & $N_\text{data}$ & $N_\text{mix}$ & MAE$^\text{2.0}_\text{mix}$ \\
    \endfirsthead

    \caption*{Table \ref{modUNI20_tab:loo_sets_hE} continued.}\\
	\toprule
    \multirow{2}{*}{$m-n$} & \multicolumn{4}{c}{mod.~UNIFAC 1.0 horizon} & \multicolumn{3}{c}{mod.~UNIFAC 2.0 only} \\ \cline{2-8}
    & $N_\text{data}$ & $N_\text{mix}$ & MAE$^\text{2.0}_\text{mix}$ & MAE$^\text{1.0}_\text{mix}$ & $N_\text{data}$ & $N_\text{mix}$ & MAE$^\text{2.0}_\text{mix}$ \\
    \midrule
	\endhead
1-2 & 11734 & 599 & 0.15 & 0.26 & 392 & 28 & 0.31\\
1-7 & 15366 & 190 & 0.55 & 0.54 & 618 & 12 & 0.36\\
1-9 & 19871 & 611 & 0.2 & 0.19 & 797 & 40 & 0.27\\
1-52 & 1664 & 35 & 0.15 & 0.13 & 3442 & 114 & 0.22\\
1-59 & 349 & 19 & 0.86 & 0.46 & 3 & 1 & 0.03\\
2-17 & 114 & 8 & 0.16 & 0.27 &  &  & \\
2-19 & 281 & 17 & 0.2 & 0.22 & 11 & 2 & 0.76\\
2-34 & 51 & 6 & 0.05 & 0.12 & 15 & 1 & 0.49\\
2-35 & 23 & 3 & 0.26 & 0.34 &  &  & \\
3-5 & 10942 & 293 & 0.36 & 0.78 & 599 & 12 & 0.55\\
3-10 & 631 & 39 & 0.27 & 1.18 & 45 & 2 & 0.12\\
3-23 & 168 & 10 & 0.28 & 0.13 & 4 & 2 & 1.41\\
3-28 & 31 & 3 & 0.09 & 0.07 & 2 & 1 & 0.18\\
4-12 & 34 & 2 & 0.01 & 0.01 &  &  & \\
4-15 & 116 & 7 & 0.09 & 0.49 & 18 & 1 & 1.65\\
4-55 & 64 & 6 & 0.11 & 0.1 & 54 & 3 & 0.18\\
4-58 & 8 & 1 & 0.03 & 0.04 & 54 & 3 & 0.18\\
5-17 & 309 & 8 & 0.37 & 0.18 & 18 & 2 & 2.25\\
5-19 & 1118 & 34 & 0.19 & 0.25 & 21 & 1 & 0.74\\
5-25 & 513 & 24 & 0.23 & 0.4 &  &  & \\
5-32 & 37 & 2 & 0.32 & 0.38 &  &  & \\
5-35 & 633 & 18 & 0.33 & 0.24 & 149 & 7 & 0.74\\
5-44 &  &  &  &  & 83 & 9 & 0.16\\
6-19 & 135 & 5 & 0.07 & 0.15 &  &  & \\
6-31 & 160 & 1 & 0.21 & 0.02 &  &  & \\
6-43 & 286 & 3 & 0.14 & 0.08 & 10 & 1 & 0.02\\
6-50 & 38 & 1 & 0.3 & 1.12 &  &  & \\
7-12 & 53 & 3 & 0.29 & 0.43 &  &  & \\
7-14 & 876 & 14 & 0.91 & 0.38 & 85 & 2 & 0.8\\
7-33 & 20 & 1 & 0.63 & 0.43 &  &  & \\
9-16 & 364 & 20 & 0.32 & 0.33 &  &  & \\
9-23 & 70 & 6 & 0.24 & 0.19 &  &  & \\
9-52 &  &  &  &  & 378 & 16 & 0.06\\
10-24 &  &  &  &  & 76 & 5 & 0.11\\
11-15 & 204 & 9 & 0.21 & 0.14 &  &  & \\
11-41 & 945 & 35 & 0.24 & 0.13 & 66 & 7 & 0.15\\
12-26 &  &  &  &  & 57 & 2 & 0.04\\
12-50 &  &  &  &  & 30 & 1 & 0.01\\
14-38 &  &  &  &  & 20 & 1 & 0.01\\
14-46 &  &  &  &  & 33 & 1 & 0.34\\
14-48 &  &  &  &  & 36 & 2 & 0.42\\
15-18 &  &  &  &  & 33 & 4 & 0.1\\
15-31 &  &  &  &  & 8 & 1 & 1.84\\
15-53 & 40 & 1 & 0.23 & 0.12 &  &  & \\
15-55 &  &  &  &  & 18 & 1 & 0.47\\
16-35 &  &  &  &  & 24 & 1 & 1.13\\
17-58 &  &  &  &  & 36 & 2 & 1.22\\
18-34 &  &  &  &  & 14 & 2 & 0.64\\
18-44 &  &  &  &  & 33 & 1 & 3.86\\
21-24 & 556 & 18 & 0.14 & 0.09 & 68 & 2 & 0.39\\
23-39 &  &  &  &  & 15 & 1 & 0.04\\
23-48 &  &  &  &  & 3 & 1 & 1.04\\
24-43 & 603 & 9 & 0.16 & 0.16 &  &  & \\
24-45 & 106 & 1 & 0.33 & 0.01 &  &  & \\
24-51 &  &  &  &  & 281 & 5 & 0.09\\
25-38 &  &  &  &  & 84 & 6 & 0.07\\
25-45 & 8 & 4 & 0.17 & 0.27 &  &  & \\
28-39 &  &  &  &  & 1 & 1 & 0.1\\
29-54 &  &  &  &  & 10 & 1 & 0.17\\
31-47 & 10 & 1 & 0.22 & 0.01 &  &  & \\
31-56 &  &  &  &  & 21 & 1 & 0.19\\
31-60 &  &  &  &  & 9 & 1 & 0.19\\
33-35 & 126 & 5 & 0.72 & 0.45 &  &  & \\
40-49 &  &  &  &  & 16 & 1 & 0.82\\
45-54 &  &  &  &  & 51 & 3 & 0.91\\
\bottomrule
\label{modUNI20_tab:loo_sets_hE}
\end{longtable}

\begin{figure}[H]
    \centering
    \includegraphics[width=0.8\textwidth]{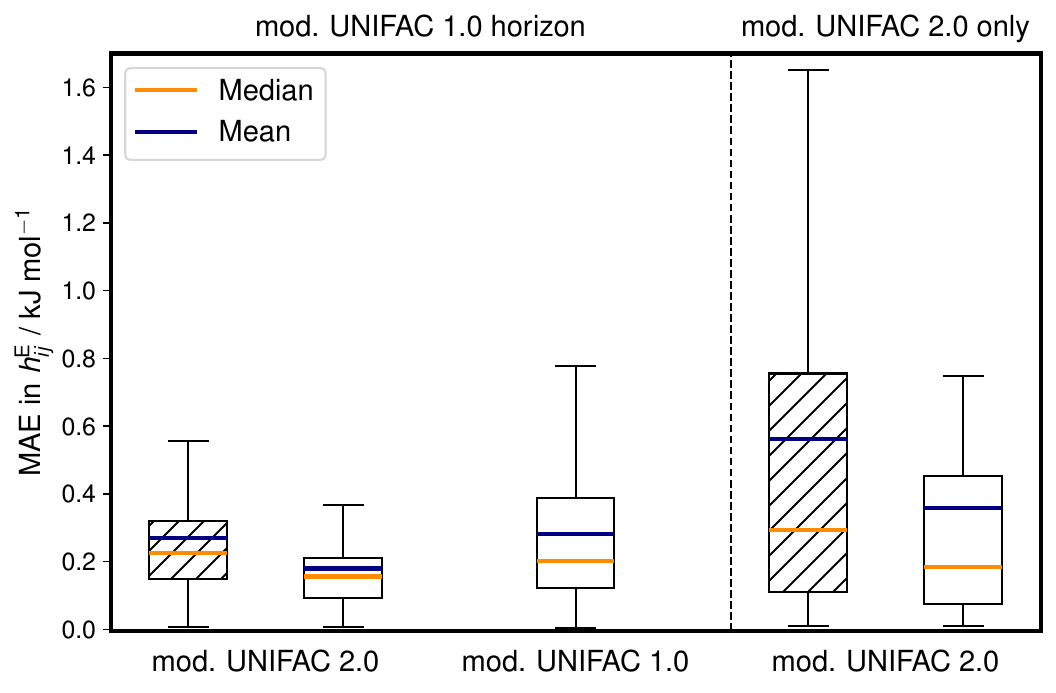}
    \caption{Mean absolute error (MAE) of the predicted $h^\text{E}$ with mod.~UNIFAC 2.0 for 100 test sets, where all data points for which a specific main group combination is relevant were withheld during training (shaded boxes); cf.~Table~\ref{modUNI20_tab:loo_sets_hE} for numerical results. The results of mod.~UNIFAC 2.0 trained on all experimental data and mod.~UNIFAC 1.0 are shown for comparison (plain boxes). The boxes represent the interquartile ranges (IQR), and the whiskers extend to the last data points within 1.5 times the IQR from the box edges.}
    \label{modUNI20_fig:Performance_hE_loo}
\end{figure}

Fig.~\ref{modUNI20_fig:Performance_hE_loo} shows the average error scores for predicting $h^\text{E}$ across all 100 test sets, analogous to Fig.~8 in the manuscript, which focuses on $\ln\gamma_i$. As before, the performance of mod.~UNIFAC 2.0 is compared to that of mod.~UNIFAC 1.0 and the version of mod.~UNIFAC 2.0 trained on the entire experimental database.
Comparing the predictions of mod.~UNIFAC 2.0 with those of mod.~UNIFAC 1.0 on the "UNIFAC 1.0 horizon" shows that the fitted pair-interaction parameters of mod.~UNIFAC 2.0 outperform those of mod.~UNIFAC 1.0, while its true predictions achieve comparable accuracy. When evaluating the true predictions of mod.~UNIFAC 2.0 against the model trained on the entire experimental data set, a slight but expected decrease in accuracy is observed. Nevertheless, the differences remain moderate, underscoring the robustness of mod.~UNIFAC 2.0 in extrapolating to unseen interaction parameters, a capability inherently lacking in mod.~UNIFAC 1.0.

\clearpage
\providecommand{\latin}[1]{#1}
\makeatletter
\providecommand{\doi}
  {\begingroup\let\do\@makeother\dospecials
  \catcode`\{=1 \catcode`\}=2 \doi@aux}
\providecommand{\doi@aux}[1]{\endgroup\texttt{#1}}
\makeatother
\providecommand*\mcitethebibliography{\thebibliography}
\csname @ifundefined\endcsname{endmcitethebibliography}  {\let\endmcitethebibliography\endthebibliography}{}